\documentclass[11pt,preprint,flushrt]{aastex}
\usepackage{graphicx,amsmath,subfigure,xcolor}

\def\eqq#1{Equation~(\ref{#1})}
\def\etal{{\it et al.}}
\def\ie{{\it i.e.}}
\def\eg{{\it e.g.}}

\newcommand{\vOmega}{\mbox{\boldmath $\Omega$}}
\newcommand{\vIstar}{\mbox{${\bf I_\star}$}}
\newcommand{\vSstar}{\mbox{${\bf S_\star}$}}
\newcommand{\vIbg}{\mbox{${\bf I_{\rm bg}}$}}
\newcommand{\vsbg}{\mbox{${\bf s_{\rm bg}}$}}
\newcommand{\vpsf}{\mbox{${\bf PSF}$}}
\newcommand{\vAeff}{\mbox{${\bf A}_{\rm eff}$}}
\newcommand{\vrr}{\mbox{${\bf r}$}}
\newcommand{\vC}{\mbox{${\bf C}$}}
\newcommand{\vH}{\mbox{${\bf H}$}}
\newcommand{\vc}{\mbox{${\bf c}$}}
\newcommand{\vRate}{\mbox{${\bf Rate}$}}
\newcommand{\vFluence}{\mbox{${\bf Fluence}$}}
\newcommand{\vRaw}{\mbox{${\bf Raw}$}}
\newcommand{\vCharge}{\mbox{${\bf Charge}$}}
\newcommand{\vGain}{\mbox{${\bf Gain}$}}
\newcommand{\vBias}{\mbox{${\bf Bias}$}}
\newcommand{\vDome}{\mbox{${\bf Dome}$}}
\newcommand{\vSFlat}{\mbox{${\bf SFlat}$}}
\newcommand{\vGRC}{\mbox{${\bf GRC}$}}
\newcommand{\vDecal}{\mbox{${\bf DECal}$}}
\newcommand{\vScience}{\mbox{${\bf Science}$}}
\newcommand{\vMask}{\mbox{${\bf Mask}$}}
\newcommand{\vWeight}{\mbox{${\bf Weight}$}}
\newcommand{\vSky}{\mbox{${\bf Sky}$}}
\newcommand{\nonlin}{\mbox{${\bf Nonlin}$}}

\newcommand{\edit}[1]{#1}

\begin{document}
\slugcomment{Version 1.4, submitted to PASP}

\title{Instrumental response model and detrending for the Dark Energy Camera}

\def\andname{}

\author{
G.~M.~Bernstein\altaffilmark{1},
T. M. C.~Abbott\altaffilmark{2},
S.~Desai\altaffilmark{3},
D.~Gruen\altaffilmark{4,5,6},
R.~A.~Gruendl\altaffilmark{7,8},
M.~D.~Johnson\altaffilmark{8},
H.~Lin\altaffilmark{9},
F.~Menanteau\altaffilmark{7,8},
E.~Morganson\altaffilmark{8},
E.~Neilsen\altaffilmark{9},
K.~Paech\altaffilmark{10,11},
A.~R.~Walker\altaffilmark{2},
W.~Wester\altaffilmark{9},
B.~Yanny\altaffilmark{9}
\\ \vspace{0.2cm} (DES Collaboration) \\
}

\altaffiltext{1}{Department of Physics and Astronomy, University of Pennsylvania, Philadelphia, PA 19104, USA}
\altaffiltext{2}{Cerro Tololo Inter-American Observatory, National Optical Astronomy Observatory, Casilla 603, La Serena, Chile}
\altaffiltext{3}{Department of Physics, IIT Hyderabad, Kandi, Telangana 502285, India}
\altaffiltext{4}{Kavli Institute for Particle Astrophysics \& Cosmology, P. O. Box 2450, Stanford University, Stanford, CA 94305, USA}
\altaffiltext{5}{SLAC National Accelerator Laboratory, Menlo Park, CA 94025, USA}
\altaffiltext{6}{Einstein Fellow}
\altaffiltext{7}{Department of Astronomy, University of Illinois, 1002 W. Green Street, Urbana, IL 61801, USA}
\altaffiltext{8}{National Center for Supercomputing Applications, 1205 West Clark St., Urbana, IL 61801, USA}
\altaffiltext{9}{Fermi National Accelerator Laboratory, P. O. Box 500, Batavia, IL 60510, USA}
\altaffiltext{10}{Excellence Cluster Universe, Boltzmannstr.\ 2, 85748 Garching, Germany}
\altaffiltext{11}{Universit\"ats-Sternwarte, Fakult\"at f\"ur Physik, Ludwig-Maximilians Universit\"at M\"unchen, Scheinerstr. 1, 81679 M\"unchen, Germany}


\begin{abstract}
We describe the model for the mapping from sky brightness to the
digital output of the Dark Energy Camera, and the algorithms adopted
by the Dark Energy Survey (DES)
for inverting this model to obtain photometric measures of celestial objects from the raw camera
output.  The calibration aims for fluxes that are uniform across the
camera field of view and across the full angular and temporal span of
the DES observations, approaching the accuracy limits set by shot
noise for the full dynamic range of DES observations. The DES pipeline incorporates
several substantive advances over standard detrending techniques,
including: principal-components-based sky and fringe subtraction; correction of
the ``brighter-fatter'' nonlinearity; use of internal consistency in
on-sky observations to disentangle the influences of quantum efficiency,
pixel-size variations, and scattered light in the dome flats; and
pixel-by-pixel characterization of 
instrument spectral response,
through combination of internal-consistency constraints with auxiliary
calibration data.  This article provides \edit{conceptual derivations of the}
detrending/calibration steps, and the procedures for obtaining the
necessary calibration data. Other publications will describe the
implementation of these concepts for the DES operational pipeline, the
detailed methods, \edit{and the validation that the techniques can
  bring DECam photometry and astrometry within $\approx 2$~mmag and
  $\approx3$~mas, respectively, of fundamental atmospheric and
  statistical limits.} The DES techniques should be
broadly applicable to wide-field imagers.
\end{abstract}

\section{Introduction}
The Dark Energy Camera (DECam) was installed at an upgraded prime
focus of the 4-meter Blanco Telescope at Cerro Tololo InterAmerican
Observatory in late 2012 \citep{decam}.  The \edit{520}-megapixel science
array tiles a 2-degree-diameter field of view (FOV) with
deep-depletion CCD detectors. 
The camera is designed for optimal imaging performance in the
$g,r,i,z,$ and $Y$ bands for the 525-night,
5000-deg$^2$ Dark Energy Survey (DES), but is also highly
productive for a broad range of astronomical investigations \citep{deswhitepaper}.  

Successful DECam science depends critically upon being able to
transform the raw output of the camera into reliable, uniformly
calibrated measures of the brightness received from astronomical
objects of interest.
The standard techniques for removal of CCD instrumental
signatures remain largely unchanged since the
earliest days of CCD astronomy \citep{GunnWestphal}: subtraction of
overscan and bias frames, division by an image of a dome screen, and
then subtraction of an estimated ``constant'' sky background.
One substantial advance was to recognize
the merit of removing background and fringing by producing a 
night-sky flat from a median of disregistered sky exposures
\citep{Tyson86}.  \edit{Examples of heavily-used array-camera detrending pipelines
include \citet{theli}, \citet{ptf}, and the \citet{desai} pipeline used for the
publicly released DES Science Verification (SV)
data.}\footnote{Available at \url{https://des.ncsa.illinois.edu/releases/sva1}}
But
these standard techniques fall well short of 
removing instrumental signatures to the floor of photometric accuracy
and reproducibility set by shot noise and stochastic atmospheric
processes.  In this paper we offer a more complete physical model of
the CCD output, which leads to general-purpose detrending algorithms
that recognize 
issues neglected by the simple procedures.  In particular, we attend
fully to the varying solid angles subtended by pixels across the
array, the impact of stray light (unfocused photons) on the
calibration process, the varying spectral response of the camera
across time and across pixels, and the proper treatment of additive
signal contaminants (background) with distinct and time-variable spectra.

\edit{These refinements of CCD detrending algorithms} are not new concepts, but to our knowledge have not been
coherently documented and treated in facility-instrument pipelines
before the advent of the current generation of FOV-filling CCD cameras
on large telescopes.  We describe our DECam response model, \ie\ the
transformations from true astronomical sky brightness into the output
pixel values, in Section~\ref{model}. Then Section~\ref{makecals}
describes the input data and algorithms for creation of all the
calibration factors entering into the DECam model.
Section~\ref{pipeline} enumerates the ``detrending'' steps in the DES Data
Management (DESDM) pipeline implements to invert the
DECam response to the sky.  We summarize the key points for users of
DECam data and precision detrending of other large general-purpose
imaging cameras in Section~\ref{conclusion}.  Table~\ref{glossary}
provides a guide to the symbols/quantities defined in this paper.  More detailed
information on aspects of the DECam detrending and calibration are
available elsewhere:
\begin{itemize}
\item An overview of DECam design, construction, and commissioning is
  in \citet{decam}.
\item \citet{desdm} describe the realization and operation of the
  DESDM pipeline, which implements the
  detrending algorithms described in this paper.  The DESDM pipeline
  executes many functions beyond single-image detrending---coaddition,
  cataloging, transient detection, and quality assessment.
\item \citet{plazas} document the ``tree-ring'' effect in the DECam
  CCD devices.
\item The ``brighter-fatter'' effect in DECam CCDs is characterized by
  \citet{gruen}. 
\item The astrometric mapping from DECam pixels to the sky is
  characterized in detail by \citet{astrometry}.
\item The principal-components sky-subtraction algorithm is detailed
  in Appendix~\ref{skypca}.
\item The ``star flat'' (Section~\ref{starflatsec})
  describing spatial and temporal photometric response variation is
  characterized in detail by \citet{starflatpaper}. The intra-night
  repeatability of DECam photometry is evaluated therein.
\item The global relative photometric calibration via joint modeling
  of the instrument and atmospheric response functions is described by
  \citet{fgcm}, including evaluation of the global photometric
  repeatability over the first three years of DES.
\item Other aspects, including absolute calibration of the DES
  data, will be addressed in future papers.
\item The proceedings of the conferences on \textit{Precision
    Astronomy with Fully Depleted CCDs} contain extensive discussion of these
  topics \citep{paccd1,paccd2}
\end{itemize}

The detrending algorithms described herein were applied to all
science-quality images taken through the SV and 
\edit{first four complete seasons of DES observation. Data from SV and the
first three season have internal designation Y3A2, and will be part of
the upcoming DR1 public data release.}
The single-exposure
images in the earlier SV public release lack
some of the refinements described herein.  These are superseded by the
Y3A2 reprocessing.  \edit{Overviews of the DES data processing
  procedures as
  envisioned before the survey are given in \citet{MohrDESDM} and
  \citet{SevillaDESDM}, and an overview of processing for the internal
  first-year processing (Y1A1) is in \citet{gold}.}

\begin{deluxetable}{cl}
\tabletypesize{\tiny}
\tablewidth{0pt}
\tablecolumns{3}
\tablecaption{Glossary of symbols}
\tablehead{
\colhead{Quantity} & 
\colhead{Description \textit{(dimensions)}}
}
\startdata
\multicolumn{2}{c}{\textit{Source quantities}} \\[2pt]
$F_\star(\lambda)$ & Spectral shape of source $(\lambda^{-1})$\\
$F_{\rm ref}(\lambda)$ & Spectral shape of reference source $(\lambda^{-1})$\\
$F_p(\lambda;p)$ & Spectral shape of source with parameter(s) $p$ $(\lambda^{-1})$\\
$f$ & Flux of source, such that $f(\lambda)=f F_\star(\lambda)$ $(E
A^{-1} t^{-1})$\\
$\vIstar(\lambda, t)$ & Surface brightness pattern of astronomical sky,
\ie\ scene incident on atmosphere $(E A^{-1} t^{-1} \Omega^{-1} \lambda^{-1})$\\
$\vIbg(\lambda,t)$ & Surface brightness pattern of background/foreground sky $(E A^{-1} t^{-1} \Omega^{-1} \lambda^{-1})$\\
${\bf S}_\star(t)$ &
Scattered light from bright sources: detected signal $(e)$ \\
$\tilde {\bf I}_{\rm ghost}(t)$  & Scattered light from bright
sources: apparent sky brightness $(E A^{-1} t^{-1} \Omega^{-1} \lambda^{-1})$ \\
${\bf s}_{\rm bg}$ & Ratio of scattered to focused
light detected from uniform background $(\cdots)$\\[2pt]

\multicolumn{2}{c}{\textit{Detrending input data, intermediate
    products, and outputs}} \\[2pt]
$ b_{tj}$ & Amplitude of background template $\vSky_j$ in exposure $t$ $(e)$\\
$\vCharge(t)$ & Charge read per pixel $(e)$\\
$\vFluence(t)$ & Source photons incident on pixel (for ref. spectrum) $(e)$
\\
$\vpsf(\lambda,t)$ & Point spread function $(\Omega^{-1})$\\
$\vRate(t)$ & Charge production rate per pixel $(e t^{-1})$ \\
$\vRaw(t)$ & Camera output values (ADU), also detrending input image $(\cdots)$\\
$\vScience(t)$ & Detrending output image, \ie\ estimator of \vFluence\
$(e)$ \\
$\vMask(t)$ & Image of data quality flags $(\cdots)$ \\
$T$ & Exposure time $(t)$\\
$\vWeight(t)$ & Inverse variance of \vScience\ attributable to read
noise and background shot noise $(e^{-2})$ \\[2pt]

\multicolumn{2}{c}{\textit{Calibration quantities and operators}}\\[2pt]
$\vAeff(\lambda,t)$ & Collecting area of telescope times detection
efficiency $(A)$ \\
${\bf BF}$ & Brighter-fatter  model \\
$\vBias(t)$ &  Line-by-line bias (overscan) \\
 BPM & Bad pixel mask \\
$\vC(t; p)$ & Flux correction in exposure $t$ for spectral
parameter(s) $p$ $(\cdots)$ \\
$\vc(t) $ & Color correction $\vC$ linearized in $p$ and converted to
magnitudes (mag per mag color) \\
$c_G(t)$ & Scalar color correction term applied to exposure $t$ (mag
per mag color) \\
$\vDecal(\lambda)$  & Narrowband-illuminated dome flat $(\cdots)$ \\
$\vDome$ & Dome flat $(\cdots)$ \\
$f_1$ & Nominal zeropoint: flux producing 1 $e$/sec/pix for
ref. spectrum $(E t^{-1} A^{-1})$\\
$\vGain$ & Amplifier gain $(e^{-1})$\\
$G(t)$ & Scalar global calibration factor for exposure $t$
$-2.5\log_{10} GRC(t)$  (mag)\\
${\bf GRC}(t)$ & Global relative calibration, \ie\
exposure-dependent portion of $\vrr(t)$ $(\cdots)$ \\
${\bf H}$ & Background subtraction operator \\
$\nonlin$ & Amplifier nonlinearity model \\
$r_{\rm ref}(\lambda)$ & Reference spectral response $(\cdots)$ \\
$\vrr(\lambda,t)$ & Spatial/temporal variation in pixel spectral response $(\cdots)$\\
$\vrr(t)$ & Pixel response variation, integrated over reference spectrum $(\cdots)$\\
$S_{\rm atm}(\lambda,t)$ & Atmospheric transmission spectrum at time
$t$ $(\cdots)$ \\
\vSFlat & Star flat, \ie\ correction from \vDome\ to $\vrr(t)$ $(\cdots)$\\
$\vSky_j$ & Background template component $j$ $(\cdots)$\\
${\bf XTalk}$ & Crosstalk model \\
$\Omega_0$ & Nominal pixel solid angle $(\Omega)$\\
\vOmega &  Deviation of pixel solid angle from nominal $(\cdots)$\\
${\bf Zero}$ & Spatial structure of bias $(e)$ 
\enddata
\vspace{-0.6cm}
\tablecomments{\textbf{Bold} quantities are images, \ie\ functions of
  pixel position. \edit{Dimensionality of quantities are given in terms of:
  energy ($E$); wavelength ($\lambda$); time ($t$); area ($A$); solid
  angle ($\Omega$); photon or photocarrier counts ($e$); ADU or
  dimensionless ($\cdots$)}}
\label{glossary}
\end{deluxetable}
 
\section{Response model}
\label{model}
\subsection{Reference bandpass and reference sources}
The purpose of astronomical imaging is to infer the surface brightness
distribution $I_\star(\theta,\phi,\lambda,t)$ of light across
celestial coordinates $(\theta,\phi)$, wavelength $\lambda$, and epoch
of observation $t$ (we will take $t$ as a discrete index
labeling exposures).  \edit{$I_\star$ has units of
  power per (area$\times$solid angle$\times$wavelength).} For photons
that are properly focused by the 
telescope, the astrometric calibration of the instrument
maps celestial coordinates into indices ${\bf x}$ of pixels on the
camera array.  We will use boldface to denote quantities that are
vectors over the array pixels, \ie\ they have an implicit discrete
argument ${\bf x}$, and arithmetic operations on these vectors are
assumed to be element-wise.
The desired sky signal is $\vIstar(\lambda,t)$.  The astrometric
solution also yields the solid angle subtended by each pixel, which we
take to be a nominal value $\Omega_0$ times a relative scaling vector
\vOmega\ over the array. We will ignore the polarization state of the
incident light and the polarization sensitivity of the instrument in
this analysis.

If the focused sky photons were the only source of signal on the
detector, then rate of production of photocarriers in the pixels
could be written as
\begin{align}
\label{rate1}
\vRate(t) & = \Omega_0 \int d\lambda\ \frac{\lambda}{hc} \vAeff(\lambda,t)
\vIstar(\lambda,t) \vOmega(\lambda,t) \\
 & = \frac{\Omega_0}{f_1\times 1\,{\rm s}} \int d\lambda \, \vIstar(\lambda,t) r_{\rm ref}(\lambda)
   \vrr(\lambda,t) \vOmega(\lambda,t).
\end{align}
Here \vAeff\ is the collecting area times the total transmission
of the atmosphere and optics, times the quantum efficiency of the detector
pixel.  In the second line we \edit{absorb the typical value of the product $\vAeff
\lambda / hc \times 1\,{\rm s}$ into
an overall constant $1/f_1.$}  A dimensionless \emph{reference bandpass}
$r_{\rm
  ref}(\lambda)$ is chosen to peak near unity and represent the
spectral response of the system for a typical pixel and observing
conditions, and the function \vrr\ whose deviation from unity
describes spatial and temporal deviations in the spectral response
from the reference or ``natural'' bandpass of the camera.  A source
with flux $f_1$ when integrated over the reference bandpass generates 1
photocarrier per second under typical conditions. \edit{The units of
  $f_1$ are power/area.}
 
For a point source, the incident surface brightness is spread over the
array by a point spread function (PSF) which we normalize such that
\begin{align}
\vIstar(\lambda,t) & = \frac{f}{\Omega_0} F_\star(\lambda) 
\vpsf(\lambda,t), \\
\sum_{\bf x} \vpsf(\lambda,t) \vOmega(\lambda,t) & = 1 \\
\int d\lambda\, F_\star(\lambda) r_{\rm ref}(\lambda) & = 1
\end{align}
Since imagers like DECam count photons without regard to their energy,
a single observation can constrain only the amplitude $f$ of the stellar
flux, not the spectral shape $F_\star(\lambda).$  \edit{The shape
  $F_\star$ has units of inverse wavelength.}

Consider first the limited goal of determining $f$ for sources which are known
to share a \emph{reference source spectrum} $F_{\rm ref}(\lambda).$  We want
to obtain the same result for $f$ regardless of the time or
focal-plane position of the observation.  To obtain $f$ if given the \vRate\ image,
we first define the \edit{dimensionless} \emph{reference flat}
\begin{equation}
\label{refflat}
\vrr(t) \equiv \int d\lambda\, F_{\rm ref}(\lambda) r_{\rm
  ref}(\lambda) \vrr(\lambda, t)
\end{equation}
and construct the \emph{fluence} image, \edit{with units of counts:}
\begin{equation}
\label{fluence}
\vFluence(t) \equiv T\, \frac{\vRate(t)}{\vrr(t)}
\end{equation}
where $T$ is the exposure time.
If we make the assumption that \vpsf\ and \vOmega\ are independent of
$\lambda$ across the filter's bandpass and the width of the PSF, we
can estimate the source flux by this sum over pixels spanning the
object's image:
\begin{align}
\label{phot1}
\left(\frac{1\,{\rm s}}{T}\right) \sum_{\bf x} \vFluence(t) & =
                                                              \frac{f}{f_1}
                                                              \sum_{\bf x}
                                         \frac{\int d\lambda F_{\rm ref}(\lambda)  r_{\rm ref}(\lambda)
                                         \vrr(\lambda, t) \vpsf(t) \vOmega(t)}
                                         {\int d\lambda F_{\rm ref}(\lambda) r_{\rm ref}(\lambda)
                                         \vrr(\lambda, t)} \\
 & = \frac{f}{f_1} \sum_{\bf x} \vpsf(t) \vOmega(t) \\
 & = \frac{f}{f_1}.
\end{align}
Thus \emph{the reference flat is the quantity we need in order to homogenize
the photometry of the survey.}  Common practice is to estimate the
reference flat by the values in an image of a source of near-uniform
surface brightness, \ie\ a \emph{flat field.}  We will describe below
why this is inaccurate---for DECam, dome flats deviate by up to
$\pm5\%$ from the reference flat---and in section~\ref{starflatsec} will
describe our method for estimating the reference flat.

Determination of \vrr\ for the entire survey would constitute
\emph{global relative calibration} as it enables accurate
determination only of the ratio of fluxes of two objects with the
reference spectrum.   A further \emph{absolute
  calibration} step is needed to determine $f_1$ if we wish to place
the fluxes on a known physical scale.

The output of the DES pipeline for exposure $t$ is an estimate of the
\vFluence\ image, which, as per \eqq{phot1}, is an estimate of the 
number of photons \emph{incident} on the pixel during the exposure for
a source with the reference spectrum.
The 
sum over pixels yields a flux (and magnitude) estimate for each
object to enter into object catalogs. The DES images and cataloged
magnitudes are hence correct only for objects sharing the reference
spectrum $F_{\rm ref}(\lambda).$
For practical reasons the reference spectrum for DES is taken to be
that of the F8IV star C26202 from the HST CalSpec standard 
set.\footnote{\url{http://www.stsci.edu/hst/observatory/crds/calspec.html}
Note that some fluxes are referenced to flat-$f_\nu$ spectra
    by \citet{fgcm}.}

\subsection{Color corrections}

Consider now the aperture photometry for a source known to
have a more general spectrum $F_p(\lambda)$ parameterized by some
value(s) $p$.  The flux estimate will be
\begin{align}
\hat f & = f_1 \times \frac{1\,{\rm s}}{T} \times \sum_{\bf x} \vFluence(t) \\
 & = f \sum_{\bf x}
   \frac{\int d\lambda F_p(\lambda)  r_{\rm ref}(\lambda)
   \vrr(\lambda, t) \vpsf(\lambda, t) \vOmega(\lambda, t)}
                                         {\int d\lambda F_{\rm ref}(\lambda) r_{\rm ref}(\lambda)
                                         \vrr(\lambda, t)} \\
\label{color1}
 & = f \times \vC(t;p), \\
\label{phot2}
\vC(t;p) & \equiv 
   \frac{\int d\lambda F_p(\lambda)  r_{\rm ref}(\lambda) \vrr(\lambda, t)}
           {\int d\lambda F_{\rm ref}(\lambda) r_{\rm ref}(\lambda)
           \vrr(\lambda, t)} 
\end{align}
The last line defines the \emph{color correction} that must be divided into
the cataloged flux in order to obtain the correct amplitude $f$ of
the source spectrum $fF_p(\lambda)$.   In obtaining \eqq{color1} we
have assumed that \vC\ is constant across the pixels that comprise the
object image, and again the \vpsf\ and \vOmega\ are independent of
wavelength across the object.

Determination of $\vC(t;p)$
most generally requires that we know not only $F_p(\lambda)$ but also
the full response $\vrr(\lambda,t)$ of the array over time,
space, and wavelength.  This general knowledge cannot be obtained
solely from internal calibrations of on-sky DECam data.

The \vC\ corrections are of course smallest and smoothest in $p$ if we have
made the sensible choice of reference response $r_{\rm
  ref}(\lambda)$ to be the system response
under typical conditions, so that $\vrr(\lambda,t)$ is near unity and
the cataloged fluxes are in the ``natural'' system of the camera.
In this case the \vC\ corrections are doubly differential: the
cataloged fluxes produced with the reference flat via \eqq{phot1} are
correct ($\vC=1$) if \emph{either} the source has the reference spectrum or the
instrument has the reference response.  The departure of $\vC$ from
unity scales as the product of the (small) variations in camera
spectral response times the (not always small) deviation of the source
spectrum from the reference. 
 
The color correction simplifies considerably if the sources are stars
with color $g-i<2$ (spectral type M0) and we assign the parameter $p\equiv
g-i-(g-i)_{\rm ref},$ where $(g-i)_{\rm ref}=0.44$ is the color of our
reference star C26202 in the AB-normalized natural bandpass of DECam.
In this case, synthetic photometry using stellar spectral
libraries \citep{ting} indicates that, for the range of $\vrr(\lambda,t)$
expected from variation in atmospheric and DES instrumental
transmission, the color correction will be fit to $<1$~mmag accuracy
by the form
\begin{equation}
-2.5\log_{10} \vC(t;p) = p \times \vc(t).
\end{equation}
Thus for the population of not-too-red stars, \emph{the photometric
  response for a given filter is fully specified by the absolute calibration $f_1$, the
  reference flat $\vrr(t)$, and the color term $\vc(t)$.}  We will show
in section~\ref{starflatsec} that the last two of these are nearly fully
constrained by internal calibration methods.

For sources with highly structured spectra, such as cool stars,
quasars, and supernovae, 
the $\vc(t)$ maps are useful constraints
on $\vrr(\lambda,t)$ but not sufficient to determine $\vC(t;p)$ to desired accuracy.
In section~\ref{colorsec} we describe how DES models the full response
$r_{\rm ref}(\lambda)\vrr(\lambda,t)$ by combining internal
comparisons of on-sky data,
narrow-band flat-field observations, and atmospheric monitoring.
This allows estimation of \vC\ for an arbitrary specified source
spectrum.  This process is described fully in \citet{fgcm}.  This
full-bandpass $\vrr(\lambda,t)$ synthesis is currently the primary method for color
calibration in DES.   The internally derived $\vc(t)$ can be used
as a crosscheck for the full synthesized $\vrr(\lambda,t).$

\subsection{Background and scattered light}

The focused photons from astronomical sources \vIstar\ comprise only
some of the light detected by the camera. For ground-based imaging,
these are outnumbered by light focused from 
\emph{background} surface brightness $\vIbg(\lambda,t).$
The distinction between signal and background is in the eyes
of the beholder: diffuse astrophysical emission from dust reflections
or intracluster light might be considered a nuisance by those
conducting photometry of more compact objects.  In
this section we will consider as
background\footnote{Foreground, for the pedantic.}
just the atmospheric emission (airglow) plus scattering of 
sunlight, moonlight, and terrestrial light in the atmosphere and from zodiacal dust.
In cloudless conditions, these sources are very smooth across the
sky.  We cannot assume they are strictly constant in a DECam exposure
since the 
distances from the zenith, the Sun, and the Moon all vary
enough across the 2-degree diameter of the DECam FOV to
produce easily-detected background gradients.
Clouds can produce more highly variable background emission and
absorption \citep{Burke}; we will discuss only clear-night processing
in this paper. 

At least several percent of the photons reaching the DECam focal plane
have arrived after unwanted reflections or scattering.  We will denote
as $\vSstar(\lambda,t)$ the pattern of photocarrier production by
photons originating in extraterrestrial sources (other than the Sun) and
arriving at the focal plane \emph{out of focus.} More precisely,
  our stellar photometry will be normalized to the flux measured
  within a 6\arcsec\ aperture. Any photons from astrophysical
  sources
whose path brings them
  to the focal plane outside such an aperture relative to their true
  direction of origin should not be counted in
  our response functions and will be assigned to \vSstar.
 There are various mechanisms producing such mis-guided photons.
  Annular
``ghosts'' appear where light from compact sources arrives after stray specular
reflections, particularly light that reflects from the focal plane and
returns after a second reflection from a filter or lens surface.
Large diffuse scattering halos also surround all sources, and photons
from sources outside the FOV can scatter from insufficiently
baffled telescope surfaces. Diffraction spikes can extend beyond our
aperture. For bright stars, these effects become
detectable and interfere with accurate measurement of focused light.
We will collectively refer to \vSstar\ as the ``ghost'' signal, and
consider it the job of the analysis software, rather than the
detrending process, to deal with these signals.  In the current DESDM
pipeline, regions of detectable ghost signals are usually just flagged as invalid.

More important to the detrending are the photocarriers
produced by light scattered from the
nearly-Lambertian background sources.  This signal can be thought of
as the integral of all the ghosts and scattered light obtained from
sources uniformly distributed across the sky.  It will be present in every
exposure, and scale with the (nearly uniform) brightness of the
background.

Our full model for the rate of photocarrier production in the camera
is hence
\begin{equation}
\label{rate2}
\vRate(t) = \frac{\Omega_0}{f_1\times 1\,{\rm s}} \int d\lambda \, \vOmega \, r_{\rm ref}(\lambda)\,
\vrr(\lambda, t) \left\{
\vIstar(\lambda, t) + {\bf \tilde I}_{\rm ghost}(\lambda,t) 
+ \vIbg(\lambda,t)\left[1 + \vsbg(\lambda, t)\right] \right\}.
\end{equation}
In this equation we've defined ${\bf \tilde I}_{\rm ghost}$ to be the
the spurious brightness signal that scattered source photons cause
in detrended images, essentially $\vSstar/\vrr.$ 
We will henceforth ignore this term as
the responsibility of the analysis software.

The term $\vsbg(\lambda,t)$ is the ratio of charge production from
scattered background light to that from focused background light at a given
detector pixel, wavelength and time.  It can be only crudely predicted from
the optics models for DES.  We need to determine this signal
empirically to subtract it from the $\vRate$ signal to desired accuracy.

The presence of background and scattered light in
the images causes two complications: first, such signals must be
removed by a sky-subtraction algorithm to isolate the \vIstar\
signals we wish to measure.  Less appreciated is that their presence invalidates
the standard method of using dome flats to estimate the reference flat $\vrr(t)$.

\subsubsection{Dome flats}
Common practice for visible and NIR image processing is to estimate
the reference flat $\vrr(t)$ by measuring the camera response to a
nearly-Lambertian source of light observed on the same date: either a
twilight sky, or the median night-sky
signal, or an illuminated screen on the dome.  We will generically 
call this the \emph{dome flat}.  If we
optimistically assume that the target source has perfectly uniform
illumination with spectrum $F_{\rm dome}(\lambda),$ \eqq{rate2} gives
the resultant signal as 
\begin{equation}
\label{dome}
\vDome(t) \propto \vOmega \int d\lambda\, r_{\rm ref}(\lambda)
\vrr(\lambda, t) F_{\rm dome}(\lambda) \left[1 +
  \vsbg(\lambda,t)\right].
\end{equation}
There are several reasons why we should \emph{not} expect this to
to yield the same pattern as the reference flat in \eqq{refflat}:
\begin{enumerate}
\item The spectrum $F_{\rm dome}$ does not match $F_{\rm ref}.$  It is
  possible to mitigate this problem by using twilight flats and
  assigning a G-star spectrum times Rayleigh scattering
    spectrum as the reference (though this would not correspond
to any real source's spectrum).  For DECam, however,
  twilight flats are prohibited since the detectors are susceptible to
  permanent damage from over-exposure.  Night-sky flats are very poor
  matches to any realistic $F_{\rm ref}$ in redder bands where airglow lines
  generate fringes because of Fabry-Perot oscillations in
  $\vrr(\lambda)$. 
\item Dome flats do not fill the telescope pupil in exactly the same
  way as distant sky sources, causing subtle differences in both the
  focused and scattered light patterns.
\item The dome flat contains a factor of pixel area \vOmega, which we
  do \emph{not} want in our reference flat if we are performing
  aperture photometry.
\item The dome flat is contaminated by the scattered light signal
  \vsbg.
\item The polarization state of the dome illumination will not be the
  same as the true atmospheric and zodiacal backgrounds, and the
  instrument may have polarization sensitivity.
\end{enumerate}
Even if the first two problems---spectral and pupil mismatches---could
be mitigated, the last three problems are still present.  It is
impossible to distinguish pixel-area fluctuations and
scattered light from true QE variations solely from an image of a Lambertian
source.  In section~\ref{starflatsec} we describe our process of
combining dome flats with internal calibration of stellar catalogs to
generate better estimates of the reference flat \vrr.

\subsubsection{Sky subtraction}
We need a \emph{sky subtraction} operator {\bf H} which we can apply
to the \vRate\ image of \eqq{rate2} that removes the background terms,
leaving us with an image obeying \eqq{rate1} that can be used
in \eqq{phot1} to measure object fluxes.  A good linear operator
would split the parts of \vRate\ that
arise from sources vs background:
\begin{align}
\vH\left[ \vRate_{\rm bg} \right] & = 0, \\
\vH\left[ \vRate_\star \right] & = \vRate_\star, \\
\vRate_{\rm bg} & \equiv \vOmega \int d\lambda\, r_{\rm ref}(\lambda,t)
               \vrr(\lambda,t) \vIbg(\lambda,t) \left[1 +
               \vsbg(\lambda,t)\right], \\
\vRate_{\star} & \equiv \vOmega \int d\lambda\, r_{\rm ref}(\lambda,t)
               \vrr(\lambda,t) \vIstar(\lambda,t).
\end{align}

The usual approach to sky subtraction is morphological, based on
the assumption that \vIbg\ varies slowly with sky position, while no
(interesting) part of the source signal does so.  A high-pass filter
or robust fitting of low-order polynomial to the image
can then be used for \vH.  But while \vIbg\ can be safely assumed to
have little spatial structure, the function $\vrr(\lambda,t)$ can have
small-scale structure due to Fabry-Perot behavior, producing fringes
in the $\vRate_{\rm bg}$ image.  Likewise the \vOmega\ function of DECam (and
most deep-depletion CCD cameras) has structure on
scales of interest in the source image due to stray electric fields,
\eg\ the ``tree rings'' analysed by \citet{plazas}. Furthermore the scattered-light function
$\vsbg(\lambda)$ can have sharp features, \eg\ when a stray reflection
path places an image of the pupil nearly at the focal plane, as
happens with the KPNO Mosaic camera \citep{mosaic}.  

A partial remedy is to divide the \vRate\ image by a dome flat
before applying the high-pass filter.  This cancels out the $\vOmega$
and $(1+\vsbg)$ terms in $\vRate_{\rm bg},$ rendering the sky smooth
again:
\begin{equation}
\label{domesky}
\vH\left[ \vRate  \right] = {\bf
  HighPass}\left[\frac{\vRate}{\vDome}\right] \times \vDome.
\end{equation}
This is imperfect, since the dome flat spectrum does not exactly match the
night-sky spectrum.  For example the dome illumination does not excite
the fringe pattern like real airglow.
A better background-subtraction scheme is to replace \vDome\ with a
median \emph{night sky flat} in \eqq{domesky}, which by definition contains the
same fringes and other structures as the $\vRate_{\rm bg}$ we are targeting.

This common approach is essentially equivalent to fitting and
subtracting a scaled version of the median sky signal from the
image. This too fails, however, if the background signal pattern
varies in time---which it does, for example, as the relative
contributions of airglow (and hence
fringes), zodiacal light, moonlight, etc. change.

We choose to dispense with the morphological approach entirely, and
instead exploit the knowledge that the background signal is determined
by a handful of physical variables, such as the phase of the moon,
the excitation state of the airglow, and the relative positions of the
telescope, zenith, moon, and Sun.  The charge collection rate should
be expressible as
\begin{align}
\label{skypca1}
\vRate(t) & = \vRate_\star + \sum_k b_{tk} \vSky_k + {\bf Noise}(t) \\
 & = ({\bf Sparse}) + ({\bf Low\,Rank}) + ({\bf Noise}),
\end{align}
where $k$ runs over a small number $K$ of sky
\emph{template} patterns.  The second line restates the
sky-subtraction problem as the standard problem of robust 
principal component analysis: the camera output, expressed as a matrix of
dimension (pixel number) $\times$ (exposure number), is expected to consist of a
sparse term (the astronomical objects, which cover a minority of the
pixels), a low-rank term (the low-dimensionality background signal),
and shot noise.  Many algorithms exist to identify the low-rank
components in this situation; we use a standard principal components
analysis (PCA) coupled with outlier rejection iteration.  Within DES we have the luxury of
thousands of exposures taken in each filter, from which we can derive
the series of sky templates $\vSky_k.$  The sky subtraction process
then consists of identifying the coefficients $b_{tk}$ of the
templates that best describe exposure $t$.  The processes for doing so
in the presence of the noise and sparse sky signal are detailed in
Appendix~\ref{skypca}.  Once we have these, the sky subtraction is
simply
\begin{equation}
\vH[\vRate] = \vRate -  \sum_k b_{tk} \vSky_k.
\end{equation}

The common
technique of subtracting a scaled median night-sky flat is equivalent
to the special case $K=1.$ 
In practice we find that 4 components are sufficient to
describe the background in nearly all clear-weather DES exposures.
The PCA method will fail if a majority of the pixels in the DES field
of view are covered by astronomical objects---though the Galactic plane
and cores of the Magellanic Clouds are likely the only sources
  visible to DECam that are large
and crowded enough to
cause this problem.  The 3--4 brightest stars in the DES footprint
cause large enough ghosts to confuse the PCA algorithm, but these
images are largely irretrievable anyway.  The morphology-based sky
subtraction algorithm of the {\sc SExtractor} code \citep{sextractor}
that runs during the cataloging phase of the DESDM pipeline removes
the sky variations from patchy clouds.

Dark current is negligible in DES images, so is not included in the model.

\subsection{Signal Chain}

\subsubsection{Brighter-fatter effect}
The detector and signal chain alter the \vRate\ values in several more
ways before producing the output digital signal.  When charge is
collected by the CCDs, it is repelled by charge that has been
collected earlier in the integration.  This has the effect of pulling
the pixel boundaries inwards to the charge, and may also increase
diffusion of incoming charge near highly-occupied pixels.  The net
effect is to shift charge so as to broaden the PSF for brighter stars,
hence the name ``brighter-fatter effect.''  A model for this effect
was proposed by \citet{antilogus}, and the parameters of this model
were fit to DECam data by \citet{gruen}.  We hence have an operator
${\bf BF}$ such that the charge actually collected in each pixel
of exposure $t$ is
\begin{equation}
\label{bf1}
\vCharge(t) = {\bf BF}\left[{\bf Rate}(t)\times T\right].
\end{equation}
\citet{gruen} also demonstrate that an algorithm that calculates and reverses the
charge shift from the Antilogus model leads to at least 10-fold
reduction of the effect both in theory and in practice, so we have a
practical inverse operator ${\bf BF}^{-1}.$

\subsubsection{Gain, nonlinearity, crosstalk, and bias}
The photocarrier count is converted to ADU by two output amplifiers
per CCD.  The process is slightly nonlinear. Fortunately the process
appears to be local, meaning that the output value for a
pixel depends only upon the charge collected in that pixel, not the
values in any other neighboring pixels.  The only exception is that
there is some crosstalk
between amplifiers, at the level of $\sim10^{-3}$ for intra-CCD amplifier
pairs and $\lesssim10^{-4}$ for inter-CCD pairs.
The values output by
the camera (the ``raw image'') are modeled as
\begin{equation}
\label{nonlin1}
\vRaw(t) = {\bf XTalk}\left[\nonlin\left[ \frac{ \vCharge(t)} { {\bf Gain} } \right] \right]+
\vBias(t).
\end{equation}
The {\bf XTalk} function takes the form of a matrix operation among
groups of pixels that are read simultaneously through the 124 CCD
amplifiers.  
We take the gain values, the nonlinearity function \nonlin\ for each amplifier, and the
{\bf XTalk} matrix elements to be fixed in time,
as there is currently no evidence for change over
time.  Temporal changes in gain can, in any case, be absorbed into
the reference flat $\vrr(t)$.

Charge transfer inefficiency (CTI) introduces non-local errors into
CCD images.  Fortunately DECam is ground-based, hence shielded by
Earth's atmosphere from cosmic rays that create charge traps in the
device.  Furthermore, DES is an exclusively broadband survey and all
science exposures are 45--330~s duration, so the
background level is high enough to render CTI unimportant.  Observers
using $u$-band or narrow-band filters with DECam, or using short
exposures, may need to revisit CTI.

\section{Constructing calibration data}
\label{makecals}
The detrending pipeline begins with the $\vRaw(t)$ images produced by
the camera and finishes with an estimate of the $\vFluence(t)$ images that
depict the flux distribution $\vOmega\times\vIstar(t)$ (assuming all 
sources have the reference spectrum).  This requires inversion of the
model defined by Equations (\ref{rate2}), (\ref{bf1}), and
(\ref{nonlin1}), as well as executing a background subtraction
operator.  The $\vC(t;p)$ function must also be determined for
inference of correct fluxes from \vFluence\ for objects with spectral
parameters $p$.
Table~\ref{glossary} lists the many calibration images and functions
that must be determined for correct detrending.
In this section we will summarize how each of these calibration
products is obtained, with full details and performance analyses of
the more complex products given in other publications.  

The
calibrations are observed to be stable on 
months-long time scales, aside from atmospheric variations and gradual
\edit{large-scale changes \vrr\ at the few-mmag level} \citep{starflatpaper}.
The camera geometry and $Y$-band response are observed to change by a few parts in
$10^3$ when the camera
temperature is cycled \citep{astrometry,starflatpaper}, and alterations
to the baffling can change the 
sky templates by 1\% or more.  We
therefore divide each DES season into \emph{observing epochs} split at
such events, and derive a single set of calibration images to be used
for a given filter during a full observing epoch.  

The dome-flat
images are observed to change by a few parts in $10^3$ from day to
day, perhaps due to subtle changes in the ambient dome light or
alignment of the dome.  The dome flats also have gradual changes of
several percent due
to aging of the illumination lamps, and some larger changes
when the illumination optics are reconfigured.  The response of the
camera to the sky is more stable than the dome flats,
which means that dividing images by daily dome flats serves to
introduce more spurious calibration variation than it removes.
We use a single \vDome\ for each epoch.  It is still advisable to take
daily dome flats, however, to monitor system performance, and be able
to notice unusual transient changes in system throughput, \eg\ from
debris on the optics.

\subsection{Bias and nonlinearity functions}
$\vBias(t)$ is determined by the standard CCD techniques: a
line-by-line subtraction of the mean overscan values is done first.
We also create a {\bf Zero} image from the robust mean of a series of
zero-length exposures, and subtract this from each image to remove
any residual spatial structure.  The {\bf Zero} signal in the DECam
CCDs is $<15e$ and smoothly structured.

The nonlinearity of each amplifier is determined from a series of
dome exposures of varying duration.  Exposure times $T$ between 0.05
and 35 seconds are selected, shuffled into random order, then
interleaved with exposures of fixed duration $T=2$~s.  The median
bias-subtracted ADU
level of pixels read through a given amplifier is calculated at each
exposure.  The 2~s exposures are used to track a small linear drift in the
dome illumination level, which is then divided out of all the
signal levels.  Nonlinearities in amplifier response are
then manifested as nonlinearities in the plot of
illumination-corrected ADU level vs exposure time.  A $\nonlin^{-1}$
function is chosen to map the ADU level back to a value proportional
to exposure time.

\begin{figure}[t]
\plotone{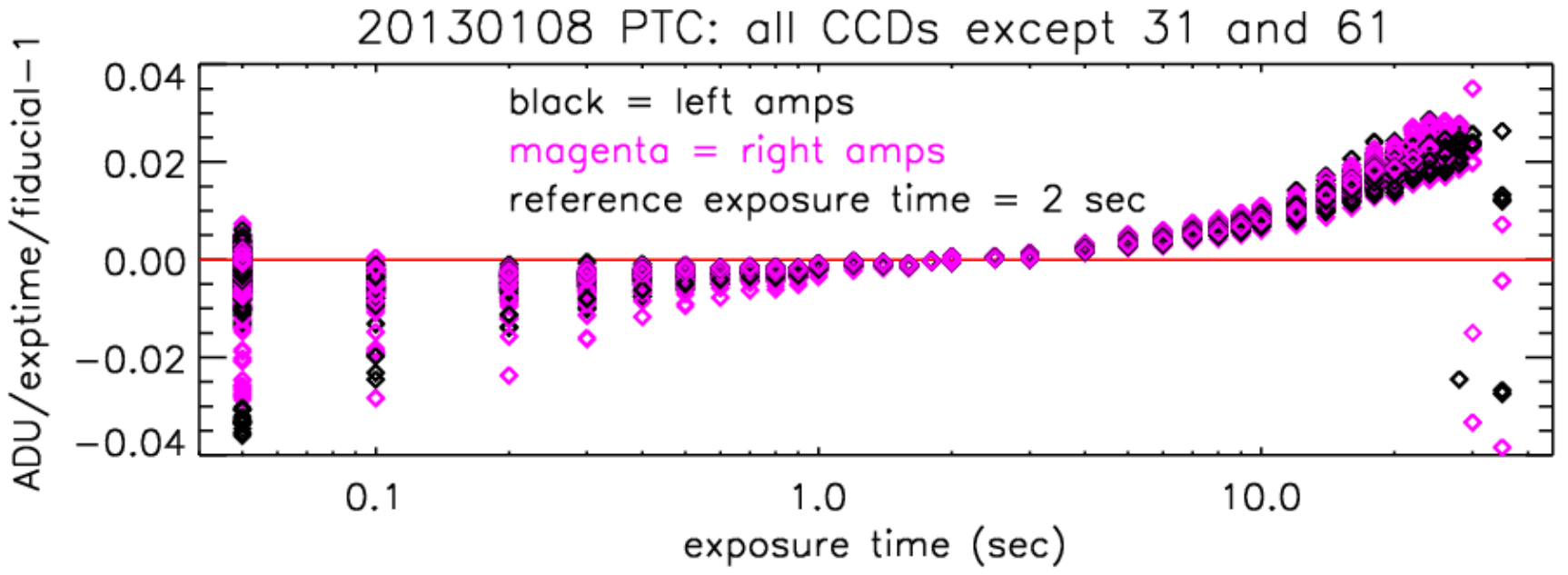}
\caption[]{Nonlinearities of 120 DECam amplifiers.  The mean signal
  level of the dome flat (after correction for lamp drift) is plotted
  against exposure time.  The $y$ axis plots the fractional change in
  ADU per second relative to exposures at $T=2$~s---perfectly linear
  devices would yield zeros.  At high light levels, all amplifiers
  show mild nonlinearity before the onset of saturation.  At low light
  levels, some amplifiers show signal deficits.}
\label{linearity}
\end{figure}
 
The DECam results in Figure~\ref{linearity} 
show two classes of nonlinearity before the onset of
saturation.  All devices show a high-light-level nonlinearity 
consistent with a quadratic response term.  A subset of devices show poorly
understood nonlinear behavior at very low illumination levels in which
some of the first few tens of electrons ``disappear.''  One 
amplifier (\#31B) shows unstable nonlinearity and is masked from all
images as unsuitable for quantitative analyses.  The worst remaining
amplifier ``loses'' $\approx20e$ of charge.  The $\nonlin^{-1}$ function
does linearize this response, though one should be cautious about
precision photometry at background levels below a few hundred $e$
because the effect is poorly understood.

\subsection{Crosstalk}
Crosstalk is nominally a linear leakage from ``source'' amplifier to
``victim'' channel, and hence there are $140^2$ leakage coefficients
to determine, since leakage may occur between any of the channels
being read out synchronously with the valid science channels
(including 5 non-science-quality amplifiers and 16 focus/alignment
channels).  Each coefficient was evaluated by plotting the median
victim amplifier output vs source amplifier signal for a large number
of sky images.  Fortunately only a handful of inter-CCD amplifiers
show detectable ($\gtrsim 10^{-5}$) crosstalk, and always at
$<2\times10^{-4}$ level.  \edit{Crosstalk between science channels and the
focus/alignment channels is small enough to ignore, so we have only the
124 science amplifiers in the crosstalk matrix.}

\begin{figure}[htb]
\plotone{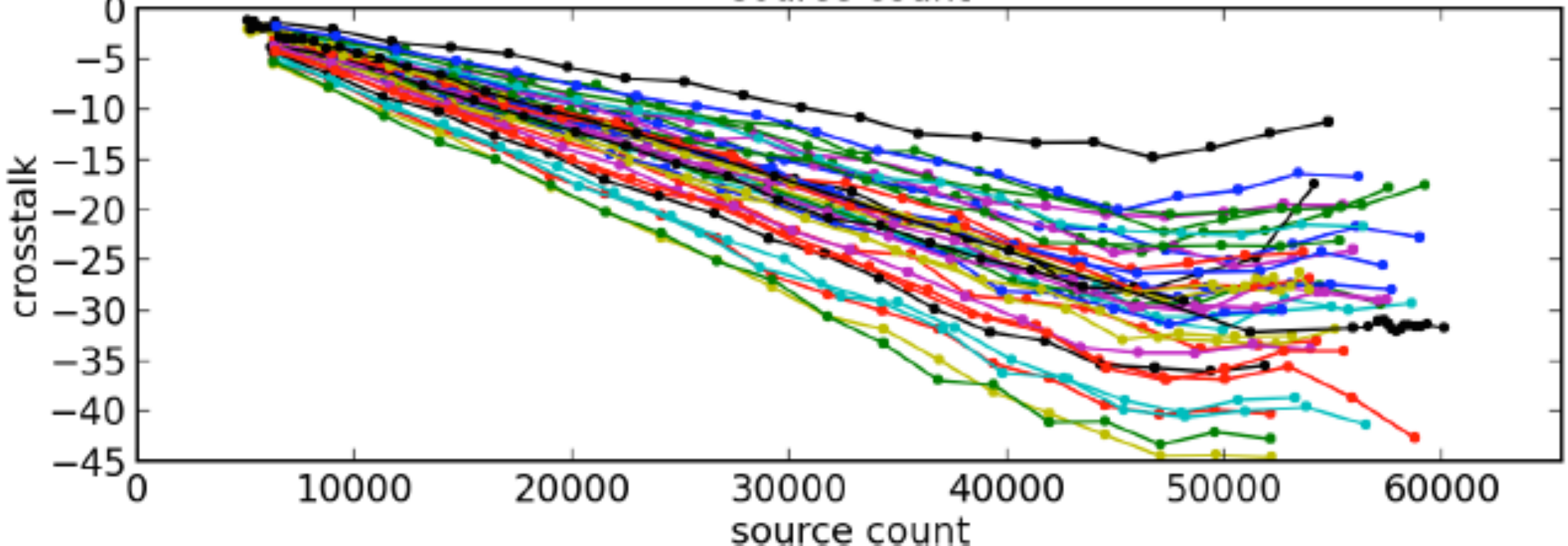}
\caption[]{Crosstalk between intra-CCD amplifier pairs.  The median
  ``victim'' signal is plotted vs source signal for pixel pairs in
  several hundred low-background sky exposures. Below the
  source saturation level, crosstalk is linear with coefficient
  $\le10^{-3}$.  Nonlinear behavior is apparent when source amplifiers
  enter their saturation regimes.}
\label{crosstalk}
\end{figure}

Figure~\ref{crosstalk} shows the victim vs.\ source trends for all of
the intra-CCD amplifier pairs, which have crosstalk coefficients up to
$10^{-3}$.  It is seen that the crosstalk becomes nonlinear when the
source amplifier exceeds its saturation level.  These nonlinearities
are tabulated and incorporated into the ${\bf XTalk}^{-1}$ operator.
Nonlinearity is not significant for the already-small inter-CCD crosstalk.

\subsection{Gain and brighter-fatter parameters}
\label{bfsec}
The model for the brighter-fatter effect, the determination of its
parameters, and the correction algorithm are fully described in
\citet{gruen}.  To summarize: the left, right, top, and bottom
($d \in \{L,R,T,B\}$) edges of pixel $ij$ on the detector are shifted
outwards by a distance determined by the charges $q$ in neighboring
pixels: 
\begin{equation}
\Delta^{d}_{ij} = \sum_{mn} a^d_{mn} q_{i+m,j+n}.
\end{equation}
The charge in pixel $ij$ is altered as
\begin{equation}
\label{bf2}
q_{ij} \rightarrow q_{ij} + \Delta^{T}_{ij} \frac{q_{ij}+q_{i,j+1}}{2}
 + \Delta^{B}_{ij} \frac{q_{ij}+q_{i,j-1}}{2}
 + \Delta^{L}_{ij} \frac{q_{ij}+q_{i-1,j}}{2}
 + \Delta^{R}_{ij} \frac{q_{ij}+q_{i+1,j}}{2}.
\end{equation}
As the $a$ coefficients are small, the effect can be reversed to good
approximation by simply changing the $+$ signs in (\ref{bf2}) to minus
signs.  The correction is fast to calculate since each $\Delta$ array is a
convolution of the {\bf Charge} image with the kernel(s) $a_{mn}$.  The
convolution is executed by fast Fourier transform; then the charge
shifting is a straightforward vector operation on the pixels.

The $a$ coefficients are the parameters of the model.  These are
determined by measuring the covariance induced by the brighter-fatter effect
between the noise on nearby pixels of dome-flat images.  These
covariances are measured to high precision from 1200 $r$-band dome
flats taken over a full season.
The covariances (and the $a$ coefficients derived from them)
are observed to be constant in time and identical for the
A and B amplifiers on a given CCD, as expected for an effect deriving
from properties of the wafer and the gate structures.

Another manifestation of the brighter-fatter effect is the suppression
of variance in dome flat images.  This invalidates the usual method of
determining the amplifier $\mathit{Gain}$, which is to note that the noise variance
$V$ expected for a pixel with expectation value $\mu$ (measured in
ADU$^2$ and ADU, respectively) should be the sum of Poisson noise and
read noise:
\begin{equation}
V = \frac{\mu}{\mathit Gain} + ({\rm RN})^2.
\end{equation}
$\mathit{Gain}$ is typically determined by regressing $V$ against $\mu$
for a series of dome flats of varying 
exposure time.  The brighter-fatter effect, however,
introduces a term $\sim a\mu^2$ into this equation.  The gain is
over-estimated if the data are fit without this term---by 5--10\% in
the case of DECam CCDs.  We determine a single gain per amplifier by a
fit that includes the brighter-fatter effect.

\subsection{Reference flat}
\label{starflatsec}
We choose to factor the reference flat image as
\begin{equation}
\label{refflat2}
\vrr(t) = \vDome \times \vSFlat \times \vGRC(t).
\end{equation}
The \emph{star flat}  \vSFlat\ is a parametric function of array
coordinates that serves to correct the errors in the dome flat.
\vSFlat\ is held constant for a given filter and observing epoch.
The time dependence of $\vrr(t)$ within an epoch is captured by 
the \emph{global relative calibration} function $\vGRC(t)$, that is
nearly constant across the field of view but can vary for each
exposure, correcting for atmospheric extinction, mirror-coating aging,
etc.  There are many fewer parameters in \vSFlat\ and \vGRC\ than
there are pixels in the array, and we can solve for these parameters
using purely internal constraints.

\subsubsection{Dome flat usage}
We have noted that the \vDome\ image is a poor approximation to the response 
$\vrr(t)$ of the array to focused light from a source with the
reference spectrum, since it confuses pixel-area variation and
scattered light with the desired measure of response to focused light.
Not surprisingly, the best way to calibrate the
camera's response to focused starlight is to measure focused
starlight, not diffuse light.  Early recognition of this principle
led to scanning a single source across the (small) arrays of the time
\citep{brian,manfroid}, but we can solve for $\vrr(t)$ much more
efficiently using the many sources detected by wide-field arrays.
If a population of objects with the
reference spectrum were spread across the sky,
we could determine the reference flat empirically
by finding a model of $\vrr(t)$ that produces uniform fluxes for all
exposures of a given reference source.  The DES photometric calibration relies primarily upon
this internal calibration method (dubbed ``ubercalibration'' by \citet{nikhil}).
Note that the internal calibration does not actually require us to know
either the reference source spectrum $F_{\rm ref}(\lambda)$ or the
reference bandpass $r_{\rm ref}(\lambda)$ in
order to determine $\vrr(t)$, nor does it offer information on
either. 

The success of this method
depends upon choosing an observation sequence that can constrain,
without degeneracy, a model for $\vrr(t)$ that adequately describes
its variations.  
Unfortunately the stellar data in DES
are insufficient to estimate \vrr\ at all 500~million DECam pixels,
never mind to do so along with time dependence.
Even with a huge number of stellar measurements, the fact that stellar
images span several pixels precludes the use of stellar data to map
response down to single-pixel scales.  Therefore we choose to rely on
the dome flats to provide small-scale response information for which a
parametric form is not readily constructed.

When the small-scale dome flat
variations are truly QE effects, this is a good idea, such as for the
``spots'' visible in Figure~\ref{starflatfig}.  But it is detrimental
in that pixel-to-pixel variations in subtended sky area, due to
imperfections in the CCD lithography, are mistakenly interpreted as QE
variations.  The latter probably lead to $\approx1$~mas RMS
astrometric errors for DECam.  It may in fact be more accurate to
replace the \vDome\ factor in Equation~(\ref{refflat2}) with a
constant \edit{except for a few regions with blemishes that are
  clearly QE features.} For future
cameras, these small-scale variations should be determined by
laboratory characterization of the devices before deployment
\citep[e.g.][for LSST]{Baumer}.

\subsubsection{Star flats}
The choices of \vSFlat\ parameterizations, and the
assignment of parameters, are described in detail in
\citet{starflatpaper}.  We give a short summary here.

The \vSFlat\ parameterization is expressed as a sum of terms in
magnitude space, \ie\ multiplicative in $\vrr.$ We begin with
independent polynomial 
variation across each CCD in the mosaic, which essentially replace the
large-scale behavior of \vDome\ with patterns constrained by stellar
data.  The \vSFlat\ model also includes terms to compensate for 
known pixel-area variations
at the edges of the devices (``glowing edges'') and in annular
patterns related to inhomogeneities in the silicon wafers (``tree
rings'') \citep{plazas}.  

The parameters of \vSFlat\ are chosen to optimize internal agreement
of stellar photometry in a series of $\approx 20$ exposures targeted
on a rich star field.  The exposures are dithered with spacings from
10\arcsec\ up to the 1\arcdeg\ radius of the DECam field, to generate
constraints on \vSFlat\ across this range of spatial scales, using
several $\times 10^5$ stellar measurements per filter.  These ``star
flat'' sequences are taken in each filter every few months during
bright time.  The \vSFlat\ for the star flat sequence is taken as the
instrumental response for the observing epoch.  For processing of
the star flat
sequences, we model $\vGRC(t)$ with a single free extinction constant per
exposure. 

The internal calibration procedure can, if applied to all stars
within the well-behaved range of color $p$, also derive the spatial and
temporal variation of the color term $\vc(t)$.  As with the reference
flat \vrr, the color term is modeled with a static function of array
coordinates, plus a per-exposure term $c_G(t)$ that is constant across the
array.  All stars in 
the well-behaved range are used to constrain \vSFlat\ and \vc\
simultaneously.  

The images are first flattened using \vDome, and fluxes for each star are
extracted using aperture photometry.
The flux $f_{\alpha t}$ of star $\alpha$ in image $t$ is
converted to $m_{\alpha t}=m_0 - 2.5\log_{10}f_{\alpha t}$ with some
nominal zeropoint $m_0$.
If $\sigma_{\alpha t}$ is the
measurement uncertainty on $m_{\alpha t}$, then we seek to minimize
\begin{equation}
\label{chi2}
\chi^2 = \sum_\alpha \sum_t \frac{\left[ m_{\alpha t} - S({\bf
      x}_{\alpha,t}) 
- p_{\alpha} c({\bf x}_{\alpha t})
- G(t) - p_{\alpha}c_G(t)
 - m_\alpha \right]^2}{\sigma_{\alpha t}^2}.
\end{equation}
We define $S=-2.5\log_{10}\vSFlat,$ and likewise the scalars
$G(t)$ and $c_G(t)$ are the logarithms of \vGRC\ and a color term
for each exposure.  The observed pixel positions ${\bf x}_{\alpha t}$
and color $p_\alpha$ of star $\alpha$ are known, and its true
magnitude $m_\alpha$ is a free parameter.  There are $O(10^3)$ free
parameters in the $S$ and $c$ functions and the $G$ and $c_G$ values.

\begin{figure}
\begin{center}
\includegraphics[height=6in]{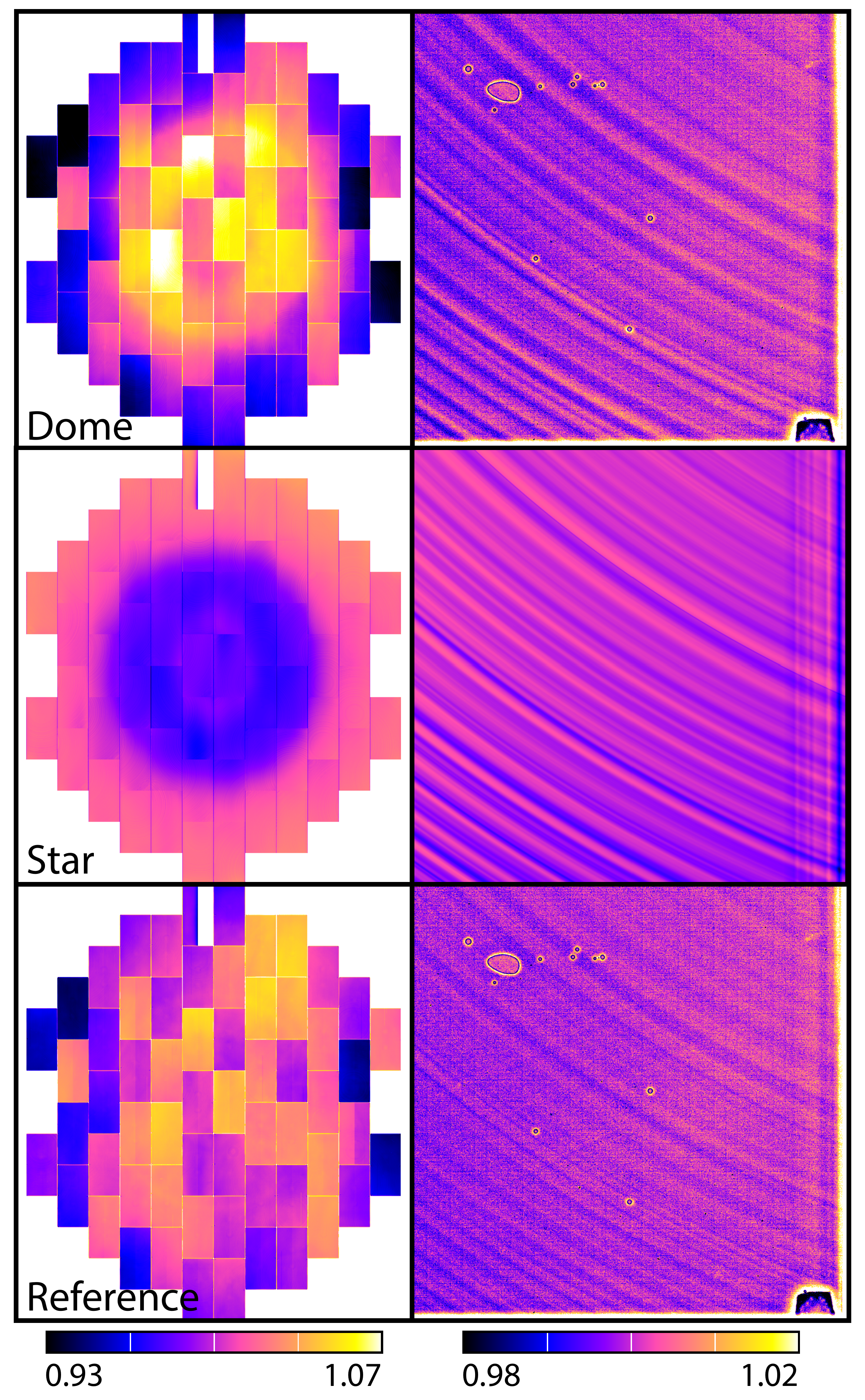}
\end{center}
\caption[]{The top row shows a dome flat from DECam through the $z$
  filter.  At left is a view of the entire focal plane, and at right
  is 1000-pixel-wide region in the corner of CCD S31 (a.k.a. CCD3).
  The middle row shows the
  parametric star flat image derived for this filter, and at bottom is
  their product, which is our best estimate of the response to focused
  starlight.
  In principle the star flat detects and removes structures in the
  dome flat due to pixel-size variation, scattered light, and color
  differences between the dome LEDs and stars, leaving only true
  efficiency variations in the reference flat.  On large scales, the
  star flat removes a scattered-light ``doughnut'' and the reference
  flat is very smooth except for CCD-to-CCD variations in QE.  On
  small scales there remain features we believe are pixel-area
  variation, but our parametric star flat models do not have the
  freedom to remove them properly.  See section~\ref{starflatsec} for details.}
\label{starflatfig}
\end{figure}

Figure~\ref{starflatfig} shows the dome flat, star flat, and their
product, the reference flat, for one of the DECam filters in one of
the observing epochs.
The second row of the Figure
  indicates that the effects measured by the star flat---scattered
  light, pixel-area variations, and color mismatches between dome
  lamps and stars---are generally smooth across the array
  and dominated by a scattered light ``doughnut'' with a peak-to-peak
  amplitude of 4\%.  The structure remaining in the reference flat is
  dominated by variations in QE from CCD to CCD, with $<0.01$~mag
  variation within each device.  The zoomed images show that most of
  the ``tree-ring'' pattern in the domes is suppressed by the star
  flat, as expected from pixel-size fluctuations, although we do not
  understand why the tree rings still remain to some extent in the
  reference flat.
  The corner ``tape bump'' features that remain in the
  reference flat are known to be from pixel-size variations, but our
  star flat parametric models do not have enough freedom to constrain
  or remove these from the reference flat.  We currently have no means
  to determine whether the remaining fine-scale features in the
  reference flat are from QE variations or pixel-size variations,
  though we suspect primarily the latter.

The optimized $\chi^2$ value is invariant under the transformations
$G\rightarrow G+G_0,$ $c_G \rightarrow c_G + c_0$,  and
$m_\alpha\rightarrow m_\alpha + G_0 + p_\alpha c_0,$ reflecting our
inability to determine the absolute magnitude or color scale with
purely internal calibrations.  External absolute calibrators are
needed to set the flux and color scales.  There are other degeneracies
in the internal calibration process; further details on the derivation
of star flats are in \citet{starflatpaper}.

\subsubsection{Global relative calibration and spectral response function}
\label{colorsec}
The above processing of specialized ``star flat'' observation
sequences determines the \vSFlat\ vector for a given epoch and
filter.  When multiplied by the dome, it yields the \vrr\ which
homogenizes the measurements of stellar fluxes across the array at the
time the star flat data were acquired.  In the DES pipeline, this
\vrr\ is used to flatten all images obtained during the epoch, and stellar
fluxes are derived from these images. 

The next step is to derive
the global relative calibration function $\vGRC(t)$ which can be
applied to these cataloged fluxes to homogenize fluxes over all $t$,
insuring that a given star would yield the same measured flux
when placed at any surveyed sky location at any time during the
survey.  Starting with Y3A1 reductions, this is done for DES
survey exposures by the \emph{forward 
global calibration module} (FGCM), described in \citet{fgcm}, which
in fact yields a model for the chromatic response function
$\vrr(\lambda, t)$ by combining a static model of the telescope/camera
spectral response with a time-varying model of the atmospheric
transmission $S_{\rm atm}(\lambda,t)$.  The atmospheric transmission
is in turn modelled with nightly values for the parameters 
for the major atmospheric optical processes: absorption by water vapor,
ozone, and well-mixed molecules; plus Rayleigh scattering and aerosol
scattering.  

The atmospheric modelling is attempted only for
cloudless nights.  For data taken through clouds, $\vGRC(t)$ is derived
by finding a distinct zeropoint shift and color term that bring each CCD of
each exposure into agreement with overlapping exposures taken on clear nights.
This procedure can benefit
from the on-site thermal-infrared all-sky imager
RASICAM \citep{rasicam}, which definitively indicates which nights are
free of clouds.

As proposed by \citet{stubbs}, FGCM splits the system spectroscopic
response into an atmospheric part (anything outside the dome)
and an instrumental function.  The atmospheric portion is taken as
constant across the field on clear nights and the instrumental
function is assumed constant for a given filter during an 
observing epoch.  Both functions are characterized using a mix of auxiliary
instrumentation and internal calibrations.

The instrumental response is measured by the DECal narrowband
flat-field illumination system \citep{decal}.  DECal directs the
output of a monochromator to the dome flat screen via optical fibers
and a projector system, which are designed to ensure that the
illumination pattern is independent of wavelength.  Furthermore the
pupil is close to uniform.  A calibrated photodiode at the top ring of
the telescope measures the brightness of dome illumination as the
monochromator is swept across each filter.

Let $\vDecal(\lambda)$ be the DECam output 
divided by the photodiode signal, so it gives the response to diffuse
light at the top ring.  If we multiply this by the
atmospheric transmission $S_{\rm atm}(\lambda,t)$, we should obtain the
response of the instrument to celestial diffuse light. Because the
DECal illuminators do not fill the telescope pupil in the exact same
way as the night sky does, the DECal signals differs from the
celestial one by a spatial
function {\bf D}.  Specializing Equation~(\ref{dome}) to a
monochromatic source spectrum, and inserting the atmospheric
transmission to get the total system response, we get
\begin{align}
S_{\rm atm}(\lambda,t) \vDecal(\lambda) & \propto r_{\rm ref}(\lambda)
\vrr(\lambda,t) \left[ 1 + \vsbg(\lambda)\right] \times \vOmega \times
                                          {\bf D}  \\
\Rightarrow \quad 
r_{\rm ref}(\lambda) \vrr(\lambda,t) & \propto S_{\rm atm}(\lambda, t) 
\times \frac{\vDecal(\lambda)}{\left[ 1 + \vsbg(\lambda)\right] \times \vOmega \times {\bf D}}.
\label{decal}
\end{align}
DECal is designed to have wavelength
dependence of {\bf D} be small, as is that of \vOmega.  If we ignore
the wavelength dependence of the scattered-light fraction
$\vsbg(\lambda)$ across the band of each filter, then the product of
\vDecal\ and $S_{\rm atm}$ give 
the spectral response $\vrr(\lambda,t)$ up to some time-independent
function of pixels.  The unknown function is determined by the
star-flat measurement of $\vrr(t).$  

The scattered-light function $\vsbg(\lambda)$ is problematic as
we have no means to distinguish scattered from focused light in the
DECal flats.  And unfortunately there is reason to suspect that it
does vary within a band: stray light from filter reflections requires (at least) one
transmission and one reflection by the filter, so
scales as $T(1-T)$ if $T$ is the filter transmission.  This signal should peak
dramatically as DECal scans across the filter shoulders.
In the future we may be
able to model this effect and refine the instrument spectral models.
Since the transition from 10\% to 90\% of peak transmission
occurs within $\le12\%$ of the bandwidth for each DES filter, we do
not expect wavelength of variation of $\vsbg(\lambda)$ to be
significantly altering the color corrections, which are already small.
 
The nightly parameters of the atmospheric transmission $S_{\rm
  atm}(\lambda,t)$ are constrained by forcing agreement on magnitudes
and colors between observations taken on different nights.  The
atmospheric constituents are further constrained by external data:
The barometric pressure and airmass for every observation are recorded
and determine the optical depths of the well-mixed molecular
components.  Synthetic photometry indicates that 
ozone variations will contribute less than 1~mmag to $\vC(t;p)$ in the
DECam bands, and hence a typical value is assumed for all observations
\citep{ting}. 
The ATMCam instrument
\citep{atmcam} continuously monitors a bright star through four
narrow filters selected to constrain the remaining
aerosol and water vapor components.  
The latter is also monitored by continuously
by a high-precision GPS receiver.\footnote{Information on measurement of
  precipitable water vapor via GPS is at \url{http://www.suominet.ucar.edu}.}  These instruments
(which are not always in service) are supplemented by internal color
constraints to estimate the remaining atmospheric values for each
night. 

With the \vDecal\ scans and the nightly atmospheric model in hand, we can
integrate any source-spectrum family to derive the spectral flux
correction factor $\vC(\lambda,t;p)$.
Recall that the star flat process yields an empirical color term
$\vc(t)$ for stellar spectra.  Stellar spectra are well known and
hence \vc\ is a known special case of $\vC(t;p)$, hence \vc\ is a
consistency check.  For example, DECal scans detect a 6~nm
center-to-edge change in the blue cutoff of the DES $i$-band filter
\citep{decal}. 
Synthetic stellar photometry predicts a radial color term that
agrees to better than 1~mmag with the values empirically measured
from the star flats \citep{ting}.

\subsection{Absolute calibration}
The absolute flux scale $f_1$ of the DES natural system in each filter
is determined through observations of spectrophotometric standard
stars. Current estimates of $f_1$ are made using the HST CalSpec
standard star C26202, which lies in one of the DES
supernova fields (which are observed $\approx$weekly) and is within the dynamic range of our imaging.
A sample of DA white dwarf stars in the DES
footprint is being measured and modelled for use as absolute calibrators.
A more thorough treatment will be presented in a later
publication. 

\subsection{Background templates}
\label{skysub}
The background subtraction technique requires derivation of the
component templates $\vSky_k.$  This is done through a
principal-components analysis (PCA) of $\approx1000$ images taken in a given
filter in a given observing epoch.  Appendix~\ref{skypca} describes
this procedure.

\begin{figure}
\plotone{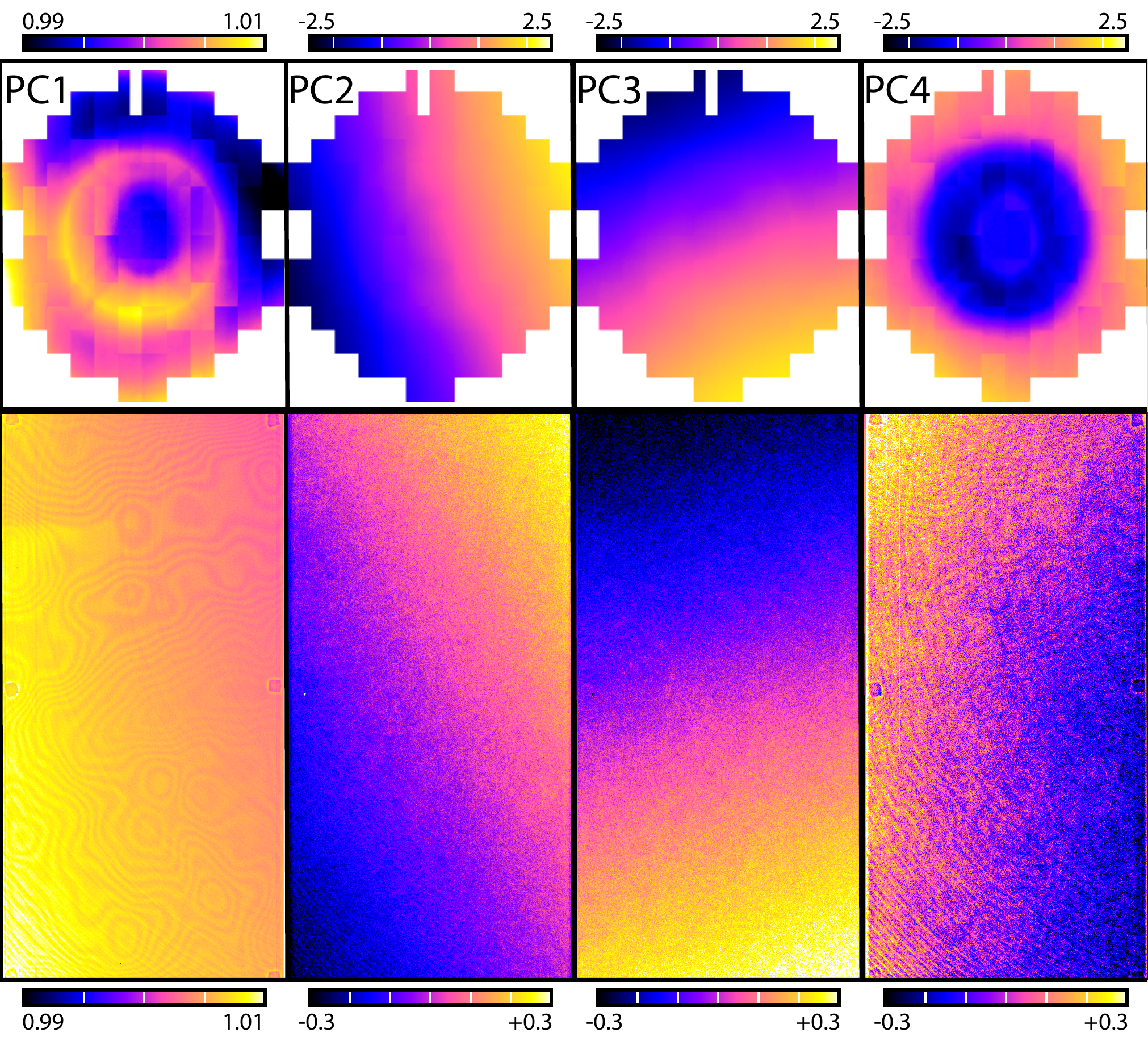}
\caption[]{The first four principal components of the background in
  the DECam $z$ filter.  At the top are subsampled views of the full
  focal plane, and at bottom are closeups of one of the edge CCDs
  (S31$=$CCD 3).  The first PC (essentially the median sky signal) shows
  structure on large scales, because the scattered-light doughnut has
  a different brightness for night-sky illumination than from the
  dome-flat illumination.  The small-scale structure is fringing.  The
  second and third PC's are nearly pure orthogonal linear gradients
  with no fringing.  The fourth PC has doughnut and fringe structure,
  consistent with a physical origin in the variation of the ratio of
  airglow to moonlight.
} 
\label{skysubfig}
\end{figure}

Figure~\ref{skysubfig} plots the four sky templates derived for the
$z$ band.  Here it is apparent that the dominant template (essentially
the median sky signal) contains signatures of fringing as well as
the scattered-light ``doughnut,'' and a color difference from the dome
flat.  Weaker components capture gradients in the sky illumination,
and changes in the ratio of airglow to continuum background.  

\subsection{Astrometric calibration}
\label{astrometry}
Determination of the map $(x,y,t)\leftrightarrow(\theta,\phi)$ between
pixel and sky coordinates is accomplished by an internal-calibration
procedure nearly completely analogous to the photometric methods,
using the same star-flat exposure sequences.
Degeneracies in the internal calibration are resolved by including
stellar positions from the Gaia DR1 catalog in the analysis
\citep{gaia}.  The full DECam astrometric model and its derivation are
described in \citet{astrometry}.

\subsection{Bad pixel mask}
For each epoch a static bad pixel mask (BPM) is created by noting pixels in
the array which have unusually high, low, or noisy levels in the bias
and/or dome flat exposures.  We also mask the 15 pixels closest to
each edge of each CCD, because distortions of the electric field near
the device edges cause changes in pixel effective area that are too
large, and potentially background-dependent, to be correctable to
the desired science accuracy.  Pixels with any of these flags set are
ignored in all processing.

The BPM additionally flags as ``suspect'' some other array regions
that are useful for most analyses, but have subtle quirks that
preclude calibration to the same accuracy as the bulk of the array.
The suspect regions include the areas 16--30 pixels from the detector
edge; the ``tape bump'' features on the CCD that have small but highly
structured pixel-size variations; and some regions that acquire excess
noise when transferring through traps or hot pixels on their way to the
readout amplifier.  Suspect pixels are detrended and carried through
to final images in the normal way, but the flag persists so that the
most precise analyses (\eg\ weak gravitational lensing, parallax, and proper
motion) can ignore these data. \citet{desdm} document the BPM flags
and the mask generating process in more detail.

\section{The detrending and calibration pipeline}
\label{pipeline}
The DES ``Final Cut'' pipeline takes the \vRaw\ images from the camera
and produces an estimate of the \vFluence\ images by inverting the photons-to-ADU model
described in Section~\ref{model}.  The output product is a FITS-format
file for each CCD containing three images: the \vScience\ image
containing the \vFluence\ estimate; a
\vMask\ bitmap image annotating the reliability of each pixel's fluence;
and a \vWeight\ image giving the inverse of the noise variance of the
pixel \emph{excluding the Poisson noise from any astronomical sources
  in the pixel.}  In this section we enumerate the operations on these
three images that comprise the pipeline.  This section provides a
mathematical description of the pipeline operations; an operational
description is in \citet{desdm}.

\begin{enumerate}
\item {\bf Bias and crosstalk removal:} A \vScience\ image is first
  created for each CCD by subtracting 
overscan and bias in the standard way, and applying the crosstalk
correction to each set of simultaneously-read pixels:
\begin{equation}
\label{step1}
\vScience(t) = {\bf XTalk}^{-1}\left[ \vRaw(t) - {\bf Overscan}(t) -
  {\bf Zero} \right].
\end{equation}

\item {\bf Linearization and gain:}
The linearization function and gain for each amplifier are applied
next:
\begin{equation}
\label{step2}
\vScience(t) \leftarrow \nonlin^{-1}\left[ \vScience(t)\right]
\times \vGain.
\end{equation}
At this point, as per \eqq{nonlin1}, the \vScience\ image should
equal the \vCharge\ image containing the number of photoelectrons
collected in each pixel.
\item {\bf Defect masking:}
Some pixels will be known at this point to have inaccurate charge
measures.  A \vMask\ bitmask image is created by combining the 
pixel defects held in the BPM file with an additional saturation flag
set for all pixels whose charge
exceeds the saturation level measured for their output amplifier.
\citet{desdm} provide a full listing of the mask bits and their meanings.

\item {\bf Brighter/fatter correction:} The operation
\begin{equation}
\label{step3}
\vScience(t) \leftarrow {\rm BF}^{-1} \left[ \vScience(t) \right]
\end{equation}
is applied using the model described in section~\ref{bfsec}.  The
\vScience\ image now should reflect the number of photoelectrons
created, rather than collected, in each pixel.  This is equivalent to
the \vRate\ image, except that we do not divide the \vScience\ image
by the exposure time $T$.

\item {\bf Dome flattening:} We execute
\begin{equation}
\label{step4}
\vScience(t) \leftarrow \frac{\vScience(t)}{\vDome}
\end{equation}
at this point, in other words we apply just one of the factors in the
reference image defined by Equation~(\ref{refflat2}).
None of the processing depends upon this being done, but applying an
approximate multiplicative correction makes it easier to inspect the
images and identify subtle signals and defects without being visually
overwhelmed by flat-field features.

\item {\bf Sky subtraction:}
The coefficients $b_{tk}$ of the background templates $\vSky_k$ are
determined as described in Section~\ref{ufit}, and the
background-subtraction operation consists of subtracting the scaled
templates:
\begin{equation}
\label{step5}
\vScience(t) \leftarrow {\bf H}\left[ \vScience(t) \right]
 = \vScience(t) - \sum_k b_{tk} \vSky_k.
\end{equation}
The sky templates are constructed from dome-flattened images and
  hence already have the dome response divided out.
\item {\bf Weight creation:}
Once we have a background estimate we can create the weight image as
the inverse of the sum of Poisson and read noise:
\begin{equation}
\label{step6}
\vWeight(t) = \frac{ \vDome^2}{ \vDome \times \sum_k b_{tk}{\bf
      Sky}_k
 + {\bf RN}^2},
\end{equation}
where {\bf RN} is the read noise (in electrons).  The \vDome\ terms
appear because the image is no longer quite in units of electrons
after the earlier step of dome flattening.  

Note that we include variance from shot noise of the
background, but do \emph{not} include shot noise in the source photons.
Analysis programs are responsible for adding the source noise term.
\item {\bf Star flattening:}  With the background now near zero, we
  apply the next term in the reference flat from \eqq{refflat2}:
\begin{align}
\label{step7}
\vScience(t) & \leftarrow  \frac{\vScience(t)}{\vSFlat} \\
\vWeight(t) & \leftarrow \vWeight(t) \times \vSFlat^2.
\end{align}
This is the output image that is saved to the DES archive.  It is not
quite equal to the $\vFluence(t)$ image that gives fluxes in the
reference system - the $\vGRC(t)$ factor from \eqq{refflat2} is missing.
$\vGRC(t)$ are applied to the 
cataloged fluxes, not to the image pixels themselves.  This is a
practical consideration: we need to produce the catalog in order to
derive the global relative calibration solutions.  It is faster to
rescale the catalog by \vGRC\ than to
re-scale the images and re-extract the sources.
\item {\bf Contaminant masking:}
Pattern-recognition algorithms are run on \vScience\ to identify pixels
that do not represent celestial fluxes; appropriate bits are set in
the \vMask\ image.  
\begin{itemize}
\item The {\sc TRAIL} bit is set for CCD bleed trails.
\item The {\sc STAR} bit is set for a circular region around each very
  bright star, extending to the radius where the stellar halo fades
  back to the sky level and will not interfere with identification of
  faint sources.
\item The {\sc EDGEBLEED} bit indicates a condition peculiar to the
  DECam CCDs: if a bright bleed trail reaches the serial register,
  charge flows along rows and obliterates a large region abutting the
  serial register for that amplifier.  An example is shown in
  Figure~\ref{edgebleed}. 
\begin{figure}[ht]
\plotone{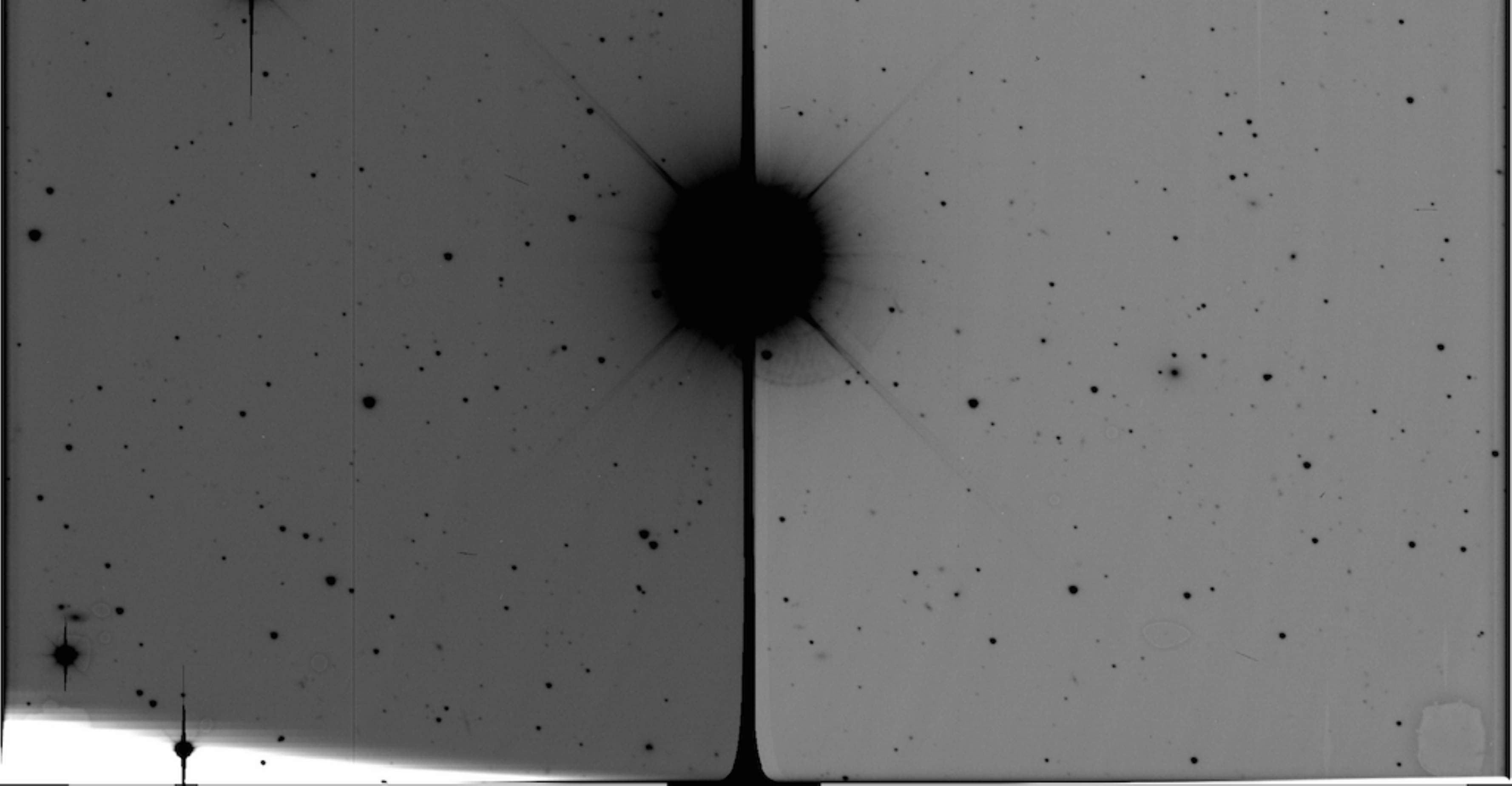}
\caption[]{An ``edge bleed'' on CCD N29 of exposure 233392, in which a bleeding
  column from a very bright star reaches the serial register, spreads
  horizontally, and perturbs the signal chain for up to an entire
  row's worth of readouts after the bleed trail.  This bleed trail
  happens to straddle the split between the two output amplifiers.}
\label{edgebleed}
\end{figure}
\item The {\sc CRAY} bit is set where cosmic rays are detected.
\item The {\sc STREAK} bit is set where meteor, asteroid, or airplane
  trails are detected.
\end{itemize}
The algorithms for identifying these features are described in
\citet{desdm}. 
\item {\bf Cataloging:}
The \vScience\ image is analyzed using the {\sc SExtractor} code
\citep{sextractor}, which produces a catalog containing fluxes,
positions, and other measurements of each detected source.  The
details of our use of {\sc SExtractor} are in \citet{desdm}.
Information from the \vMask\ and \vWeight\ images are also passed to
{\sc SExtractor} so that it can calculate uncertainties and
appropriately flag objects containing invalid or suspect pixels.
This catalog and the images then become the basis of DES science
analyses.  The weak gravitational lensing pipeline, for example,
returns to the \vScience\ images to perform more exacting measures of
galaxy shapes than {\sc SExtractor} does.
\item {\bf Global calibration:}
The cataloged fluxes of high-$S/N$ stellar sources are fed to the
global relative calibration algorithms \citep{fgcm} to derive the functions
$\vGRC(t)$ (including color terms) that homogenize the full survey
magnitudes.  As the survey progresses we will recalculate the global
solution multiple times with multiple methods. Methods used in
previous DES processings/releases are described in \citet{Tucker07}
and \citet{gold}.
When objects are retrieved from the catalog database,
their fluxes/magnitudes can be adjusted using a chosen global
solution. Chromatic corrections can be tabulated given an estimate of the
source spectral shape $F_\star(\lambda),$ or approximated using the
source color and the linearized color corrections $\vc(t)$.
\item {\bf Astrometric calibration:}
The final astrometric calibration, like the global photometric
solution, is done at catalog level.  The
parameters of the global astrometric solution(s) are used to register
the images for co-addition as well as for science analyses.  Good
solutions are derived using \textsc{scamp}, and more detailed
solutions (including chromatic terms) are available from the methods
described in \citet{astrometry}.
\item {\bf Absolute calibration:}
An overall magnitude zeropoint determination will be the last step in the
calibration process.  This requires reduction of exposures targeted on
spectrophotometric standards, which then must be tied into the global
relative calibration of the survey.  The resultant magnitude zeropoint
from such an analysis can be applied to the full catalog by the object
database. Interim zeropoints are in use from synthetic photometry of
  the HST CalSpec star C26202.

\end{enumerate}

\section{Conclusion}
\label{conclusion}

We have provided a coherent mathematical model for the output images
of DECam in terms of the astronomical brightness distribution, and a
series of algorithms to invert the process to determinations of object
fluxes.  This careful audit shows that dome flats alone are insufficient to
determine the response of the array to stars of a given flux: we use on-sky
observations of stellar sources to infer ``star
flat'' corrections for scattered light and pixel-area variations
in the dome flats.  These corrections exceed
0.04~mag for DECam.  \citet{starflatpaper} demonstrate that these
procedures homogenize the photometry to $\approx2$~mmag accuracy
across time periods of hours, \edit{and that changes in DECam photometric response
are slow (weeks to months) and limited to $\approx 5$ mmag, with only
low-order variation across the focal plane.}

We also present a PCA-based method for removing  
zodiacal light, airglow, and atmospheric scattered light signals from
the images.   Pixel-area variations and fringing are among the effects
that can impart small-scale structure on the detector output from
these ``backgrounds,'' even though they are highly uniform on the sky.
This confounds morphological algorithms for distinguishing background from
astrophysical signals of interest.  Our algorithm instead relies on
the repeatability of the background signals to distinguish them.

There are several subtleties in the interpretation of astronomical
images from DECam (and other imagers) that one must attend to for
precision analyses:  
\begin{itemize}
\item Are the values in the images representing the fluence in the
  pixels (total incident photons), or the mean surface brightness in
  the image?  These differ by a factor of pixel area.
  Aperture magnitude measurements (summations) assume the
  former, while model-fitting measurements usually assume the latter.
  The DESDM pipeline produces fluence images.
\item The pixel centers do not form a square grid on the sky when
  pixel areas vary.
  Model-fitting algorithms and morphological measures must be aware of
  this even if the images are in surface-brightness normalization.
\item The response $\vrr(\lambda, t)$ of the instrument to focused light depends on array
  position, wavelength, and time.  In producing a single calibrated
  output image for an exposure, we need to designate some nominal
  source spectral shape $F_{\rm ref}(\lambda)$ to define the
  reference flat field $\vrr(t)$ that is applied to exposure $t$.  The
  image is only homogeneously calibrated for sources with this
  spectrum; a position- and time-dependent correction must be applied
  to homogenize the calibration for sources with different spectral
  shape.  The DES catalogs now model and tabulate these chromatic corrections.
\item The PCA sky subtraction algorithm only attempts to remove light
  from near-Lambertian background sources that repeat in every exposure with only a
  few degrees of freedom---namely airglow, zodiacal light, and
  atmospheric scattered light.  Episodic contaminants
  such as stray reflections from bright stars are identified with
  different algorithms.  Downstream codes must make the division
  between background and signal for diffuse sources such as Galactic
  dust reflection or intra-cluster light.
\item Likewise the weight maps in DESDM outputs present the (inverse)
  variance only from backgrounds and read noise.  The contributions
  from shot noise in sources should be estimated from a source model,
  not from the counts in the image.
\end{itemize}

The model and inversion algorithms we construct for DECam should be
generally applicable to wide-field astronomical
imagers.  Indeed there are commonalities between the DES pipeline and
those being used for other current wide-field CCD surveys such as
PanSTARRS1 \citep{magnier}, the Kilo-Degree Survey \citep{kids}, and
the Hyper-Suprime Cam survey \citep{HSC}.  DECam
can be calibrated very reliably (as quantified in other papers)
because the optics and detector are very stable and well-behaved.  The
detectors have very little ``personality,'' with only mild
non-linearities and no known significant hysteresis.  The edge bleed
phenomenon is the only substantial quirk that affects our
high-background exposures.  The calibration images,
gains, etc., are found to be stable for months at a time, \edit{apart from
slow, low-order drifts in photometric response.}
Furthermore DECam is
on the equatorially mounted Blanco telescope, thus the camera and
telescope form a rigid unit.  Newer telescopes
are exclusively alt-az mounts, so there is continuous rotation between
telescope and camera, which may substantially complicate the behavior
of the calibrations.

The code implementing these detrending algorithms is all
  publicly available as part of the DESDM software repository.  Most
  of the detrending steps (2--8 in Section~\ref{pipeline}), as well as
  the derivation of the PCA sky templates, are implemented in the
  \texttt{pixcorrect} Python package and would require minimal
  alteration, if any, for use on data not adhering to DECam formats.
  All detrending steps are efficiently implemented as \texttt{numpy} array
  operations.

\acknowledgements

GMB gratefully acknowledges support from grants AST-1311924 and AST-1615555
from the National Science Foundation, and DE-SC0007901 from the Department
of Energy. Support for DG was provided by
NASA through the Einstein Fellowship Program, grant PF5-160138.

Funding for the DES Projects has been provided by the U.S. Department
of Energy, the U.S. National Science Foundation, the Ministry of
Science and Education of Spain, the Science and Technology Facilities
Council of the United Kingdom, the Higher Education Funding Council
for England, the National Center for Supercomputing Applications at
the University of Illinois at Urbana-Champaign, the Kavli Institute of
Cosmological Physics at the University of Chicago,  the Center for
Cosmology and Astro-Particle Physics at the Ohio State University, the
Mitchell Institute for Fundamental Physics and Astronomy at Texas A\&M
University, Financiadora de Estudos e Projetos,  Funda{\c c}{\~a}o
Carlos Chagas Filho de Amparo {\`a} Pesquisa do Estado do Rio de
Janeiro, Conselho Nacional de Desenvolvimento Cient{\'i}fico e
Tecnol{\'o}gico and the Minist{\'e}rio da Ci{\^e}ncia, Tecnologia e
Inova{\c c}{\~a}o, the Deutsche Forschungsgemeinschaft and the
Collaborating Institutions in the Dark Energy Survey.  

The Collaborating Institutions are Argonne National Laboratory, the
University of California at Santa Cruz, the University of Cambridge,
Centro de Investigaciones Energ{\'e}ticas,  Medioambientales y
Tecnol{\'o}gicas-Madrid, the University of Chicago, University College
London, the DES-Brazil Consortium, the University of Edinburgh, the
Eidgen{\"o}ssische Technische Hochschule (ETH) Z{\"u}rich,  Fermi
National Accelerator Laboratory, the University of Illinois at
Urbana-Champaign, the Institut de Ci{\`e}ncies de l'Espai (IEEC/CSIC),
the Institut de F{\'i}sica d'Altes Energies, Lawrence Berkeley
National Laboratory, the Ludwig-Maximilians Universit{\"a}t
M{\"u}nchen and the associated Excellence Cluster Universe,  the
University of Michigan, the National Optical Astronomy Observatory,
the University of Nottingham, The Ohio State University, the
University of Pennsylvania, the University of Portsmouth,  SLAC
National Accelerator Laboratory, Stanford University, the University
of Sussex, Texas A\&M University, and the OzDES Membership
Consortium. 

The DES data management system is supported by the National Science
Foundation under Grant Number AST-1138766. The DES participants from
Spanish institutions are partially supported by MINECO under grants
AYA2015-71825, ESP2015-88861, FPA2015-68048, SEV-2012-0234,
SEV-2012-0249, and MDM-2015-0509, some of which include ERDF funds
from the European Union. IFAE is partially funded by the CERCA program
of the Generalitat de Catalunya.

\appendix
\section{Principle-component sky subtraction}
\label{skypca}
We start from the assumption in Equation~(\ref{skypca1}) that any
given count-rate image $\vRate(t)$ is the sum of a sparse component
(the astrophysical sources of interest), a zero-mean noise component,
and a background component that is a linear function of a small number
$K$ of sky ``templates.''  For this Appendix we will use the notation
that ${\bf R}_t, {\bf S}_t,$ and ${\bf N}_t$ are the count rate, sparse
(object) component, and noise components of exposure $t$, expressed as vectors
over the $N_{\rm pix}$ pixel positions.  We define matrices $R, S,$
and $N$ such that element $t,i$ of each is the value at pixel $i$ in
exposure $1\le t \le N_{\rm exp}$.  We thus seek the $N_{\rm pix}\times K$ matrix $V$ of
sky template vectors and the $N_{\rm exp}\times K$ matrix $U$ of
coefficients per exposure such that
\begin{equation}
\label{SUN1}
R = S + U V^T + N.
\end{equation}
This ``robust principle components'' problem is common, \eg\ in
computer vision applications, and the subject of  much algorithm development
\citep{candes}. But it is quite unmanageable in its native form, since
the matrix $R$ is of dimensions $\approx 10^3\times 10^9$  for the set
of exposures from which we wish to build the template set.  We will
therefore create a decimation operator $D$ which compresses an image
vector ${\bf R}$ to a feature vector $\tilde {\bf R}$ of length $N_f\ll N_{\rm pix}.$
We will choose $D$ to be linear in the sense that
\begin{equation}
D(a_1 {\bf R}_1 + a_2 {\bf R}_2) = a_1 D({\bf R_1}) + a_2 D({\bf
  R}_2),
\end{equation}
and that it does not destroy the sparsity, \ie\ $\tilde {\bf S}=D({\bf
  S})$ is also a sparse in the sense of having the majority of its
elements be free of source flux.  In this case, we can apply $D$ to
each row (image) of the $R, N, V,$ and $S$ matrices and transform
(\ref{SUN1}) to
\begin{equation}
\label{SUN2}
\tilde R = \tilde S + U \tilde V^T + \tilde N,
\end{equation}
with the $U$ vector conserved.  As long as $D({\bf
  V}_k)\ne 0,$ meaning that the decimation does not project away (or
greatly attenuate) the signal of background template ${\bf V}_k,$ we
can as a first step solve for $U$ on the reduced-size problem
(\ref{SUN2}).  Then in step (2), 
we can obtain the full-resolution templates $V$ with a column-by-column
solution to Equation~(\ref{SUN1}).  Step (3) will be to derive
coefficients $u_k$ to apply to an arbitrary new image ${\bf R}$ that
was not part of the $N_{\rm exp}$ images used to construct the
templates.  

\subsection{Deriving coefficients $U$}
Our operator $D$ should yield a decimated $\tilde {\bf R}$ which
is altered by all of the physical parameters expected to
alter the background.  We expect changes in the large-scale pattern of
focused and scattered background light to arise from changes in lunar
phase and the positions of the sun, moon, and telescope.  We divide
the image into $N_D\times N_D$ pixel subarrays, and produce as $\tilde {\bf
  R}$ the medians of each subarray.   Typically we take $N_D=128.$  The resultant ``mini-sky''
image is not only smaller, but has greatly reduced noise and is more
sparse than the original image.  Objects smaller than $N_D$ pixels are
filtered away by the median, and large disturbances to the background
(such as ghosts of bright stars, reflection nebulae, or intra-cluster
light) occupy the same fraction of the compressed image's pixels as in
the original.

A concern is that airglow fringing, one of the most important backgrounds to
remove, will be largely suppressed by the median at scale $N_D,$
such that PCA of $\tilde R$ cannot recover the elements of $U$ that
depend on the strength and excitation state of the airglow.  We
therefore take an $8\times8$ block median of
one of the CCDs (which still resolves the fringes), and append to the
data vector $\tilde {\bf R}$ a $2\times2$ subsampling of this image.
The final feature vector $\tilde R$ has $N_f\approx10^{4.5}$ elements
and the linear algebra operations are much faster than the disk access times
for the exposures.

With a compressed $\tilde R$ that should manifest all changes to the
background, we proceed to execute a robust PCA decomposition via the
following algorithm:
\begin{enumerate}
\item Begin a rank-one approximation of $\tilde R \approx {\bf u} {\bf v}^T$
  with vectors ${\bf u}$ and ${\bf v}$ by initializing $v_k=1.$
\item Set $u_t = \textrm{median}_i (\tilde R_{ti}/v_i).$ \label{ustep}
\item Set $v_i = \textrm{median}_t (\tilde R_{ti}/u_t).$  Iterate with
  step~\ref{ustep} to convergence of ${\bf u}$ and ${\bf v}.$
\item Rescale $\tilde R_{ti} \leftarrow \tilde R_{ti}/(u_tv_i)$ so all
  elements are near unity.
\item Conduct a standard PCA of $\tilde R$ and retain the
  $N_{\rm keep}\sim 20$ most significant components in a trial
  decomposition $\tilde R \approx U\tilde V^T.$ \label{pcastep}
\item For each exposure $t$, define $\sigma_t$ as the RMS value of the
  residual mini-image $\tilde {\bf R}_t - \sum_k U_{tk} \tilde {\bf V}_k.$ 
\item For some chosen $\sigma$-clipping threshold $g$, replace all
  pixels having residual $>g\sigma$ with the value of the
 rank-$N_{\rm keep}$ approximation.  This step excludes super-pixels that remain
  contaminated by sources and ghosts.  Exposures with very large $\sigma$
  are discarded from the process as ill-described by the low-rank
  decomposition. 
\item Iterate to step~\ref{pcastep} until $U$ converges,
  typically 3--4 iterations.
\item Rescale $U$ and $\tilde V$ with ${\bf u}$ and ${\bf v}$ to
  make them represent the low-rank decomposition of the original
  $\tilde R$ matrix.  We further rescale $U$ and $\tilde V$ such that
  the elements of each row of $\tilde V$ (which are decimated sky templates)
  has RMS value of unity.  The $U_{tk}$ elements then give the typical
  amplitude (in count rate) of the contribution of template $k$ to the
  background of exposure $t$.
\item Select the number $K$ of templates to retain in the sky model
  by noting where the amplitudes $U_{tk}$ of the corresponding
  corrections become insignificant.
\end{enumerate}

\subsection{Deriving full-resolution templates $V$}
With $U$ in hand, Equation~(\ref{SUN1}) separates into an independent
robust linear regression for each pixel $i$ with unknowns $V_{ki}$:
\begin{equation}
R_{ti} = S_{ti} + \sum_{k=1}^K U_{tk} V_{ki} + N_{ti}.
\end{equation}
We solve this system for each $i$ using a standard linear regression
with aggressive $\sigma$-clipping of outliers, since we expect a
substantial fraction of the $N_{\rm exp}\approx10^3$ samples at this
pixel to be ``contaminated'' by object flux.  This is the most time
consuming portion of the analysis since there are $5\times10^8$ such
solutions to execute per DECam filter.  But this operation is
trivially parallelized and the computation time is insignificant
compared to the DESDM pipeline execution time.

\subsection{Fitting coefficients to exposures}
\label{ufit}
Once the sky template matrix $V$ is known for a given filter during a
given epoch, we can execute the sky subtraction on any exposure's data
${\bf R}_t,$
whether or not it was among the $N_{\rm exp}$ used to derive the
templates.  This is again a robust linear regression operation, which
we conduct on a decimated version of the exposure using the decimated
templates $\tilde V$:
\begin{equation}
\left[D({\bf R}_t)\right]_i = \tilde S_t + \sum_{k=1}^K u_{tk} \tilde
V_{ik} + \tilde N_{ti}.
\end{equation}
We solve for the $K$ coefficients $u_{tk}$ which best satisfy this
equation, once again using a $\sigma$-clipping iteration to exclude
super-pixels that are perturbed by large-scale objects or image
artifacts.  

This fit to the mini-sky image yields valuable diagnostic information
as well as the background coefficients $u_{tk}$: 
the fraction $f_{\rm clip}$ of
super-pixels that were clipped away; the RMS variation $\sigma$ of
those super-pixels that remained; and the residual of the mini-sky
image to the best sky model.
A high $f_{\rm clip}$ usually indicates
the presence of a large spurious-light source in the image, such as an
airplane trail, or ghosts of a bright star.  A high $\sigma$ value
indicates an aberrant background pattern, such as can occur for images
taken through patchy clouds, or with a very bright, nearby moon.
These artifacts are readily identified by visual inspection of the
mini-sky residuals.

\subsection{Results for DECam}

For DECam, we find that
$>99\%$ of the background signal is
described as a linear combination of $K \le 4$ sky components (except
for exposures taken through patchy clouds).  Figure~\ref{skysubfig}
shows the first four principle components of background in the DECam
$z$ filter.  The top row shows the ``mini-sky'' decimated images
 for each component row of $\tilde V$.
The second row shows a closeup of a single CCD
to reveal the presence of fringing signals in the full-resolution
$V$.  All of the signals here are shown relative to the dome flat pattern.

The dominant PC1 is basically the median night-sky signal.  The upper
left panel reveals that the background signal differs from the dome flat
in several respects: first, there is a significant slope, suggesting
that there is a gradient in the dome illumination system; second, the
donut structure suggests that the scattered light amplitude differs
between dome and sky illumination; and we also see CCD-to-CCD step
functions, which are likely due to a spectral differences between dome
and night sky combined with varying spectral response of the CCDs.  
The lower left panel reveals the fringing signal, which we expect to
be present in the night sky but not in the domes.  In the bluer
filters there is less fringe signal, but one can see in the sky
templates other small-scale deviations between dome and sky signals,
which are likely attributable to color differences.  A
morphology-based sky subtraction would fail to identify such issues.

In all the filters we find that PC2 and PC3 are essentially two
orthogonal linear slopes of sky flux across the array.  These are
readily associated with gradients in brightness expected across the
FOV as the distance to the moon or sun varies, or gradients in the
zodiacal brightness.  Note there are no fringes visible in PC2 or PC3:
the airglow is not involved in these gradients.  PC2 and PC3
coefficients are found to be usefully diagnostic of problems such as
the occultation of the telescope beam by the observatory dome.

Fringes reappear in PC4 for $z$ band, as does a donut pattern.  Our
hypothesis is that PC4 is generated by a change in the ratio of
airglow intensity to the intensity of backgrounds that have nearly-solar
continuum spectra (twilight,
moonlight, and zodiacal light).  PC4 basically allows the fringe
amplitude to vary independently of the large-scale patterns, but we see
from the upper-right panel that this is accompanied by a change in the
overall illumination pattern.  Apparently the airglow/solar ratio
change is manifested as an overall color shift of the night sky as
well as in the fringing patterns.  In fact we have found the
large-scale sky pattern is
fully predictive of the fringe amplitude for this filter: we recover the fringe
pattern in PC4 even if we omit from our decimation operator $D$ the
secondary small-scale feature vector described above.  In practice, therefore, we have
found the DECam PCA can be conducted just with the $128\times128$
median-filtered ``mini-sky'' images.  

Principal components at $k>4$ are found to contain smooth quadratic and
increasingly higher-order polynomial
variation across the focal plane, with no visible fringe or small-scale
components.  As such patterns are at most a few $e$ amplitude in
normal images, and are
getting into the realm where they may be excited by astrophysical
sources such as reflection nebulae or bright star ghosts, we truncate
our templates at $K=4$ and leave higher-order sky variation to the sky
subtraction algorithm executed at the cataloging step.

In the $g, r,$ and $i$ bands the fringes are nearly absent, but other
small-scale structures (such as tree rings) are visible in PC1.
The DECam detectors have low fringe amplitudes because their thick,
deep-depletion bulk and good anti-reflection coatings yield high
quantum efficiency to the silicon bandgap cutoff.  This makes them
weak Fabry-Perot resonators for the airglow lines.  Other detectors with
stronger fringe patterns may benefit even more than DECam from this
PCA-based sky subtraction technique, because their fringe patterns
will be stronger and may also exhibit detectable pattern shifts as excitation conditions
of the ionospheric radicals change.

\end{document}